\documentclass[journal]{IEEEtran}
\usepackage[pdftex]{graphicx}
\usepackage{cite}
\usepackage{amsmath}
\usepackage{amsfonts}
\usepackage{ascmac}
\usepackage{algorithm}
\usepackage{algorithmicx}
\usepackage{algpseudocode}
\usepackage{subfigure}
\usepackage{amsmath}
\usepackage{amsfonts}
\usepackage{amsmath,amssymb}
\usepackage{bm}
\usepackage{multirow}
\usepackage{makecell}
\usepackage{mathtools}
\usepackage{textcomp}
\usepackage{url}
\usepackage[utf8]{inputenc}
\usepackage{stackengine}
\usepackage{multirow}
\usepackage{txfonts}
\usepackage[normalem]{ulem}
\usepackage{bbding}

\newcommand{\shintani}{\textcolor{black}}
\newcommand{\sntn}{\textcolor{black}}
\newcommand{\isaka}{\textcolor{black}}
\newcommand{\isakaFix}{\textcolor{black}}
\newcommand{\isakaNew}{\textcolor{black}}
\newcommand{\isa}{\textcolor{black}}
\newcommand{\proof}{\textcolor{black}}

\usepackage{tikz}

\renewcommand{\baselinestretch}{1.0}

\algnewcommand\algorithmicforeach{\textbf{for each}}
\algdef{S}[FOR]{ForEach}[1]{\algorithmicforeach\ #1\ \algorithmicdo}

\ifCLASSINFOpdf

\else
 
\fi

\begin{document}

\title{Systematic Unsupervised Recycled \proof{Field-Programmable Gate Array} Detection}

\author{Yuya~Isaka,
        Michihiro~Shintani,
        Foisal~Ahmed,
        and~Michiko~Inoue
\thanks{Y.~Isaka and M.~Inoue are with Nara Institute of Science and Technology, Ikoma, Japan. (e-mail:\{isaka.yuya.iw6,kounoe\}@is.naist.jp).}
\thanks{M.~Shintani* is with Kyoto Institute of Technology, Kyoto, Japan. (e-mail:shintani@kit.ac.jp).  Corresponding author.}
\thanks{F.~Ahmed is with the Computer Systems Department, Tallinn University of Technology,
  Tallinn, Estonia.}
}


\maketitle

\begin{abstract}
With the expansion of the semiconductor supply chain, the use of recycled
field-programmable gate arrays (FPGAs) has become a serious concern.
\sntn{Several methods for detecting recycled FPGAs by analyzing the
ring oscillator (RO) frequencies have been proposed; however, most
assume the known fresh FPGAs (KFFs) as the training data
in machine-learning-based classification.}
{\sntn{In this study, we propose a novel
recycled FPGA detection method based on an unsupervised anomaly
detection scheme when there are few or no KFFs available.}}
As the RO frequencies in
the neighboring logic blocks on an FPGA are similar because of
systematic process variation, our method compares the RO frequencies
and does not require KFFs.
The proposed method efficiently identifies recycled FPGAs through
outlier detection using direct density ratio estimation. Experiments
using Xilinx Artix-7 FPGAs demonstrate that the proposed method
successfully distinguishes recycled FPGAs from 35 fresh FPGAs. In
contrast, a conventional
recycled FPGA detection method results in certain
misclassification.
\end{abstract}


\IEEEpeerreviewmaketitle

\IEEEpeerreviewmaketitle
\section{Introduction}\label{sec:intro}
\IEEEPARstart{W}{ITH}
the rapid progress of integrated circuit (IC) supply chain
globalization, the distribution of recycled ICs in the market as new
products has become a serious concern~\cite{IEEEProc2014_Guin,JETTA2014_Guin}.
Among the various ICs, field-programmable gate array (FPGA) is
\proof{of particular interest for} recycling because of the increase in
distribution
owing to the growing demand for deep-learning
applications~\cite{FPT2016_Nurvitadhi,DAC2016_Wang,TCAD2017_Wang}.
Manufacturers cannot guarantee
their reliability
due to aging
degradation induced by
previous usage. \proof{Therefore, recycled FPGAs not only} result in
economic loss but also cause serious problems, particularly
when used in mission-critical applications such as automobiles and
medical equipment.

Various recycled FPGA detection methods have been
proposed~\cite{DFT2014_Dogan,ICCD2017_Jyothi,ITC-Asia2019_Ahmed,TCAD_Ahmed,TVLSI2019_Alam}. The
general idea behind these methods involves the analysis of the
aging-induced degradation of circuit characteristics using the ring
oscillator (RO) frequency. The RO frequencies of fresh units are
measured and used to train a machine learning model. As these
frequencies degrade with usage, the trained model can classify the
FPGA under test (FUT) as fresh or recycled. These
methods~\cite{DFT2014_Dogan,ICCD2017_Jyothi,ITC-Asia2019_Ahmed,TCAD_Ahmed}
assume the presence of known fresh FPGAs (KFFs).
{\sntn{For accurate classification through machine learning algorithms, FPGA manufacturers need to measure a large volume of KFFs. If there are less KFFs available, these methods may not work properly.}}

To meet this requirement, an
unsupervised recycled FPGA detection method \proof{was} proposed
in~\cite{TVLSI2019_Alam}.
However, because this method uses the measured frequencies as
the input vector for the $k$-means++ clustering
algorithm~\cite{k-meanspp} as is, its classification accuracy remains
limited because of process variation.
Besides, the characterized logic blocks for RO measurement must be
carefully selected because the classification accuracy is \isaka{poor} if
they are inappropriate and/or insufficient for the measurement.


\begin{figure}[t]
  \centering
  \includegraphics[width=0.57\linewidth]{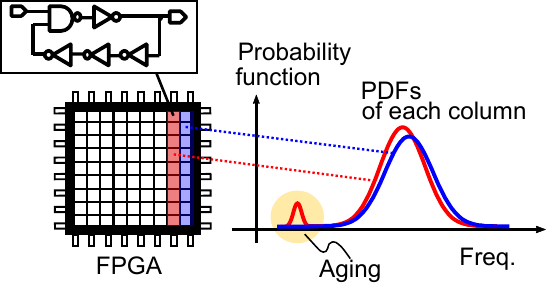}
  \caption{Key concept \proof{illustration} of the proposed method. \proof{RO frequencies} on
    neighboring logic blocks are similar because of the systematic
    component of the process variation. By comparing these
    frequencies, the proposed method can highlight the aging effect on
    ROs, realizing unsupervised recycled FPGA detection.}
  \label{fig:concept}
\end{figure}

Herein, we propose a novel method for detecting recycled FPGAs \proof{independent} of KFF by comparing the frequencies on neighboring logic blocks.
The key concept of the proposed method is shown in
Fig.~\ref{fig:concept}.  When designing ROs in each logic block and
calculating the probability density functions (PDFs) of these
frequencies in each column, the frequency distributions of neighboring
columns ideally match because of the systematic component of the
process variation~\cite{TSM2004_Ohkawa,TED2008_Saxena}. However, this
assumption does not hold if there is degradation due to aging on one
side.
The proposed method removes the need for KFFs by exhaustively
comparing all the neighboring ROs. Based on the similarity of the
compared frequencies, recycled FPGAs are detected through direct
density ratio estimation~\cite{ICML2007_Bickel} formulated as
unsupervised anomaly detection.  \sntn{We would like to note that the
  proposed method not only has the advantage of not requiring KFFs but
  also can complement the classification results of the conventional
  supervised methods when KFFs are available.}

This manuscript is based on our previous
work~\cite{IOLTS2021_Isaka}.  While our previous evaluation
insufficiently uses 10 FPGAs with a small benchmark circuit under a
single aging scenario, a more practical evaluation is provided to show
the effectiveness of the proposed method by increasing the number of
FPGAs to 35 and using RISC-V~\cite{riscv} under various aging
scenarios.
{\sntn{
The 35 FPGAs used in experiments were manufactured in different lots
and the experiments using them show the applicability of the proposed
method to FPGAs having the various initial process variations are
different.  RISC-V has been very attracting attention in recent years,
and many implementations on FPGA have been reported. By using the
processor, we evaluate the proposed method under the more practical
and possible scenario. The various aging scenarios contain a short
aging time, and the experimental results clarify the performance limit
of the proposed method experimentally, that is how long the
recycled/aging time on FPGA can be detected by the proposed method.}}


The main contributions of this paper are summarized as:
\begin{itemize}
  \setlength{\itemsep}{0pt}
\item Taking advantage of the fact that ROs in adjacent blocks have
  similar frequencies, we formulate an efficient detection scheme
  using an unsupervised outlier detection method through
  self-referencing based on density ratio estimation.
\item Silicon measurement results using 35 commercial FPGAs
  \proof{demonstrate the proposed method successfully detected the} aged
  FPGAs, excluding FPGAs with very short aging time, \proof{and could successfully} \isaka{classify}
  \proof{fresh FPGAs}. In contrast, the conventional
  method~\cite{TVLSI2019_Alam} \proof{fails to detect the aged FPGAs.}
\end{itemize}

The remainder of this paper is structured as follows.  In
Section~\ref{sec:related}, we review an existing unsupervised
recycled FPGA detection method.
In addition, we explain the direct density ratio estimation, which is
used as an unsupervised outlier detection algorithm in the proposed
method. Section~\ref{sec:propose} describes the proposed
recycled FPGA detection method. The experimental procedure and silicon
measurement results using commercial FPGAs are discussed in
Section~\ref{sec:exp}. Finally, \proof{we conclude} our paper in
Section~\ref{sec:conclusion}.

\section{Preliminaries}\label{sec:related}
\subsection{Conventional method}
In~\cite{TVLSI2019_Alam}, \proof{an}
unsupervised
FPGA detection method has been proposed\proof{,
which we briefly review}
as the conventional method. This method involves two steps: RO
frequency measurement and outlier detection based on \isaka{an} unsupervised machine learning algorithm using the
measured frequencies. If the FPGA has been previously used, the RO frequencies
will degrade because of aging mechanisms, such as bias temperature
instability (BTI)~\cite{CICC2006_Bhardwaj}, hot carrier injection
(HCI)~\cite{IOLTS2009_Lorenz}, and time-dependent dielectric breakdown
(TDDB)~\cite{TDSC2008_Srinivasan}.  Suspicious FPGAs thus can be
detected through unsupervised outlier detection by capturing the aging effect.

\begin{figure}[t]
  \centering
  \includegraphics[width=0.62\linewidth]{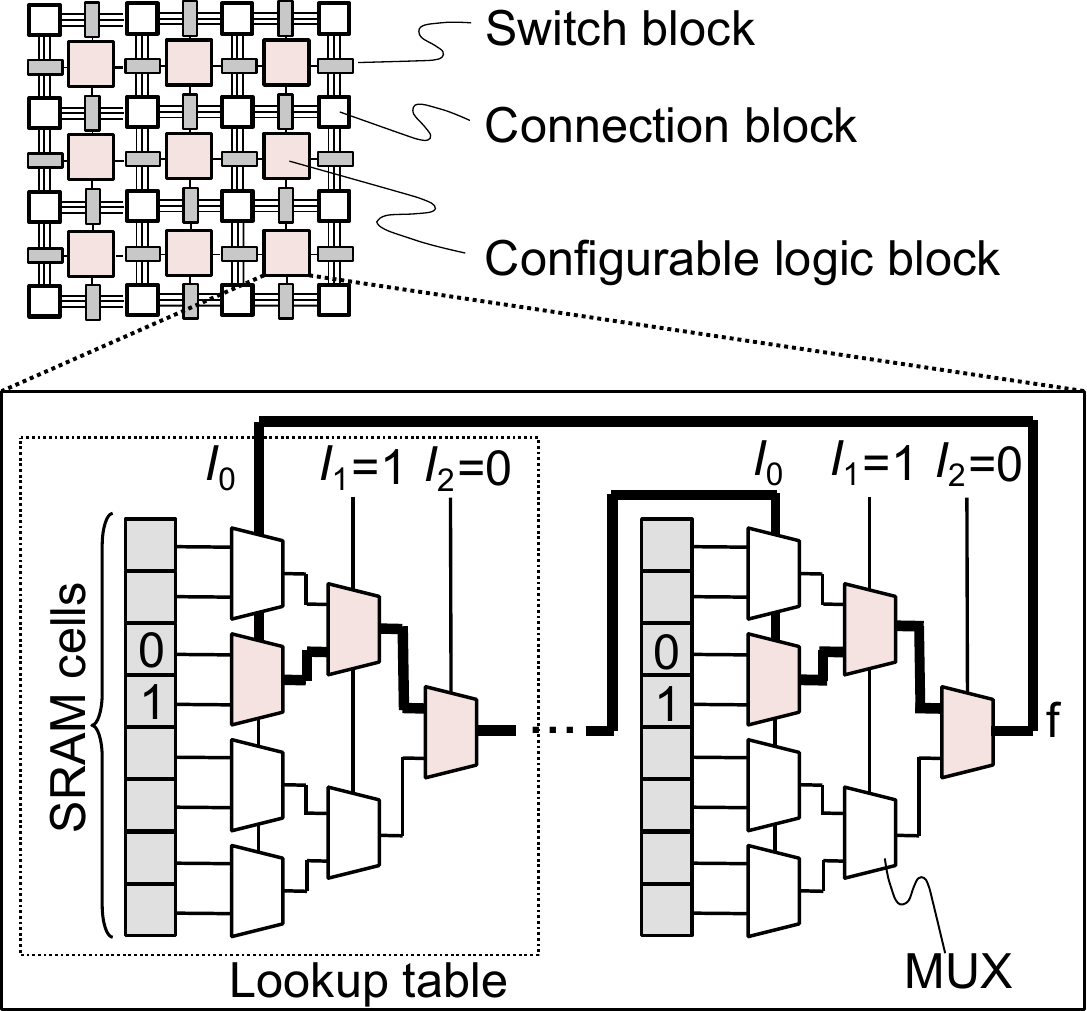}
  \caption{Typical FPGA structure. A multistage RO can be designed in a
    single CLB using multiple LUTs. By setting $I_1=1$ and $I_2=0$,
    the SRAM values are propagated, and the highlighted \isakaFix{LUT} path
    oscillates. The XP measurement enables the characterization of 
    all the LUT paths~\cite{TVLSI2019_Alam}.}
  \label{fig:structure}
\end{figure}



The general structure of an FPGA is depicted in
Fig.~\ref{fig:structure}. It includes several lookup tables (LUTs)
in one configurable logic block (CLB), which can be used to design a
multistage RO within a single CLB. In the design, as one LUT corresponds
to one stage, an odd-number-stage RO
can be configured.

In~\cite{TVLSI2019_Alam}, an exhaustive LUT path measurement
(hereafter called XP) has been proposed, enabling the analysis of all
the LUT paths by appropriately setting the LUT input value and using
XOR and XNOR gates for the RO stage.
For a $z$-input LUT,
$2^{z-1}$ paths
can be configured.
In Fig.~\ref{fig:structure},
a three-input LUT
($I_0$, $I_1$, and $I_2$) is used as an example,
where four ($=2^{3-1}$)
paths can be analyzed.  After the measurement, the FUTs are tested using
an unsupervised outlier detection method based on the measured
frequencies. In~\cite{TVLSI2019_Alam}, $k$-means++ clustering is
exploited to partition the measured frequencies. If the FUT
has been recycled, the number of clusters
is more \proof{than} that of a
fresh one with a
distance.
To quantitatively evaluate the distance, the silhouette
value~\cite{JCAM1987_Rousseeuw} is introduced in~\cite{TVLSI2019_Alam}.

All the \isakaFix{LUT} paths in the CLB are completely covered in XP measurement,
whereas the tested CLBs \proof{are} carefully selected for the
successful application of $k$-means++ clustering.
More importantly, because
the raw values of the measured frequencies that will be affected by
process variation are used as the input vector for $k$-means++
clustering,
this method may fail to detect recycled FPGAs.


\subsection{Anomaly detection via direct density ratio estimation}\label{sec:ulsif}

In this subsection, we
explain the core idea of a direct density ratio
estimation~\cite{KIS2011_Hido}, \proof{as} used in our proposed method.
This process is a well-established algorithm for robust outlier
detection, and it is performed based on the \proof{non-similarity} of the two
vectors with $n$ elements,
$\bm{f}=\{f_1, f_2, \cdots f_i, \cdots, f_n\}$ and $\bm{f'}=\{f'_1,
f'_2, \cdots f'_j, \cdots, f'_n\}$, which are the inlier data and
tested data (i.e., suspicious data), respectively. Although direct
density ratio estimation can work even when the numbers of elements in
$\bm{f}$ and $\bm{f'}$ are unmatched, the number of elements in both
vectors is set to $n$ for the sake of simplicity in this paper. By
evaluating the data as
vector, the noise in each sample can be
suppressed, and the systematic features in the vectors can be
effectively clarified.
Furthermore, the detection sensitivity is
considerably higher \proof{than} the comparison of each element (i.e.,
residual), even when the probability of outliers in $\bm{f'}$ is
small.

In the direct density ratio estimation, the \proof{non-similarity} is modeled as
the anomaly score $a(\cdot)$ which is represented as the log-arithmetic of
the density ratio
as follows:
\begin{equation}
  a(f) = -\log r(f),
  \label{eq:density_ratio}
\end{equation}
where $r(\cdot)$ is the density ratio function and is represented
as:
\begin{equation}
  \isaka{r(f)\equiv\frac{p'(f)}{p(f)}.}
  \label{eq:objective_function2}
\end{equation}
Here, $p(\cdot)$ and $p'(\cdot)$ are the probability density functions
of
$\bm{f}$ and $\bm{f'}$, respectively. When $f$ is an outlier, $a(f)$
\proof{goes higher}.

It is necessary to obtain the density ratio function
$r(f)$ based on the given vectors as shown in
Eq.~(\ref{eq:objective_function2}).
Because the direct solution of
$r(\cdot)$ involves a division operation, the error
can be large. To
address this issue, several methods have been proposed to estimate
$r(\cdot)$ without obtaining the probability distribution of
$p(\cdot)$ and $p'(\cdot)$, such as Kullback-Leibler importance
estimation procedure (KLIEP)~\cite{AISM2008_Sugiyama} and
the least-squares importance fitting (LSIF)~\cite{JMLR2009_Kanamori}. In
\proof{our proposed detection method}, $r(\cdot)$ is estimated using the unconstrained
LSIF (uLSIF)~\cite{KIS2011_Hido}, which achieves the best performance
among the density ratio estimation methods.

In the uLSIF, the density ratio function is approximated as a linear
\proof{importance model}:
\begin{equation}
  \hat{r}(f) = \sum_{l=1}^n \alpha_{l}
  \varphi_{l}(f),
  \label{eq:linear_model}
\end{equation}
where $\alpha_{l}$ and
$\varphi_{l}(f)$ are the fitting parameter and non-negative basis
function, respectively. Here, we consider the radial basis function
(RBF) kernel as the basis function:
\begin{equation}
  \varphi_{l}(f) = K(f,\isaka{f_{l}'})=\exp \left(-\frac{||f-\isaka{f_{l}'}||\isaka{^2}}{2w^2} \right),
  \label{eq:rbf}
\end{equation}
where $w$ is the bandwidth. The fitting parameters, $\bm{\alpha}$
$(={\alpha_1, \alpha_2, \cdots \alpha_n})$, are calculated so as to
minimize the squared error of $r(f)$ and $\hat{r}(f)$,
$J_0$, which is
expressed as
\begin{align}
  J_0(\bm{\alpha})&:=\frac{1}{2}\int(\hat{r}(f)- r(f))^2 p(f) df \nonumber \\ 
  &=\frac{1}{2}\int\hat{r}(f)^2 p(f)df - \int\hat{r}(f) p'(f) df \nonumber \\ 
  & \,\,\,\,\, + \isaka{\frac{1}{2}\int r(f)^2 p(f)df}.
\label{eq:eq1}
\end{align}
In Eq.~(\ref{eq:eq1}), the third term can be
ignored because it is
constant. Through
empirical approximation, an optimization problem can be formulated as
follows:
\begin{equation}
  \min_{\{\alpha_{l}\}_{l=1}^n}\left[\frac{1}{2} \sum_{l,l^\prime=1}^n\alpha_l \alpha_{l^\prime}\hat{H}_{l,l^\prime}
  - \sum_{l=1}^n\alpha_l\hat{h}_{l}
  + \frac{\lambda}{2}\sum_{l=1}^n\alpha_{l}^2\right], \label{eq:optimization}
\end{equation}
where
\begin{align}
  \hat{H}_{l,l^\prime}&:=\frac{1}{n} \sum_{i=1}^n K(\isaka{f_i},\isaka{f_{l}'}) K(\isaka{f_i},\isaka{f_{l'}'})  \,\,\,\,{\rm and}\\
  \hat{h}_{l}&:=\frac{1}{n} \sum_{j=1}^n K(\isaka{f_j},\isaka{f_{l}'}).
\end{align}
In Eq.~(\ref{eq:optimization}),
$\frac{\lambda}{2}\sum_{l=1}^{n}\alpha_{l}^2$ is a
regularization term. The solution $\tilde{\bm{\alpha}}$ is analytically given
by
\begin{equation}
  \tilde{\bm{\alpha}} = (\tilde{\alpha}_1,\tilde{\alpha}_2,\cdots,\tilde{\alpha}_n)^\top = (\hat{H} + \lambda \bm{I})^{-1}\bm{\hat{h}},
  \label{eq:solution}
\end{equation}
where $\bm{I}$ and $\bm{\hat{h}}$ are an $n$-dimensional identity matrix and a vector
of $\hat{h}_{l}$, respectively. Note that, since elements of
$\tilde{\bm{\alpha}}$ can \isaka{take} negative values, it
must be modified as
\begin{equation}
  \hat{\alpha}_{l} = \max(0,\tilde{\alpha}_{l}) \,\,\,{\rm for}\,\,\,l=1,2,\cdots,n.
\end{equation}
Thus, the probability density ratio $r(\cdot)$ can be
directly estimated, and the anomaly score vector $\bm{a}=\{a(f_1),
a(f_2), \cdots, a(f_n)\}$ for each element of $\bm{f}$ and $\bm{f'}$
can be \proof{calculated using} Eq.~(\ref{eq:density_ratio}).
Further, binary classification is performed using the anomaly score.

The primary advantage of the uLSIF over the KLIEP and LSIF is that the
fitting parameters $w$ and $\lambda$ can be analytically calculated
through leave-one-out cross-validation (LOOCV) using
Eq.~(\ref{eq:solution}).

\section{Proposed method}\label{sec:propose}
We propose a novel unsupervised
method for detecting recycled FPGAs to
realize better detection accuracy. The proposed method also utilizes
the measured RO frequencies and classifies FUTs as either recycled or
fresh. The distinction from the conventional KFF-free method is that
the necessity of the KFF set is removed by comparing neighboring
frequencies to effectively classify the FUTs.
In addition, the RO analysis is exhaustively carried out for all the
LUT paths of all the CLBs to avoid omission using a technique, called
called X-FP measurement and proposed in~\cite{TCAD_Ahmed}.  For
classification, self-referencing outlier detection is formulated based
on the uLSIF algorithm.

{\sntn{
The proposed method is based on an unsupervised anomaly detection
framework utilizing the uLSIF introduced in
Section~\ref{sec:ulsif}. In contrast to the supervised methods such
as~\cite{DFT2014_Dogan,TCAD_Ahmed}, the proposed method requires no
golden samples and works well when the amount of golden samples is
either limited or nonexistent. In addition, we would like to note
that the proposed method is applicable to various commercial
FPGAs. The proposed method assumes that the target FPGAs have the
typical structure shown in Fig.~\ref{fig:structure}. Although the
proposed method also requires a design environment in which ROs can be
configured for each LUT in order to measure frequency fingerprint as
in~\cite{TCAD_Ahmed}, it can be realized by using an open-source
design environment described in~\cite{MICRO2020_Murray}.  }}

\subsection{Key idea}
For manufactured ICs, the process variation can be decomposed into
random and systematic components~\cite{TSM2004_Ohkawa,TED2008_Saxena}.
The former is caused by physical phenomena such as random dopant
fluctuation and is
often modeled as a normal distribution with zero mean. \proof{Meanwhile},
the latter is a gradual spatial variation in the die and
modeled as a polynomial of the die coordinates $(x,y)$, where $x$ and
$y$ are the position coordinate on an FPGA.
The systematic components of the neighboring locations are
similar because the frequencies of the neighboring ROs are
ideally matched, which suggests that comparing the frequencies of
adjacent CLBs also helps in eliminating the need for KFFs.


\subsection{Procedure}
\begin{figure}[t]
    \centering
    \includegraphics[width=0.53\linewidth]{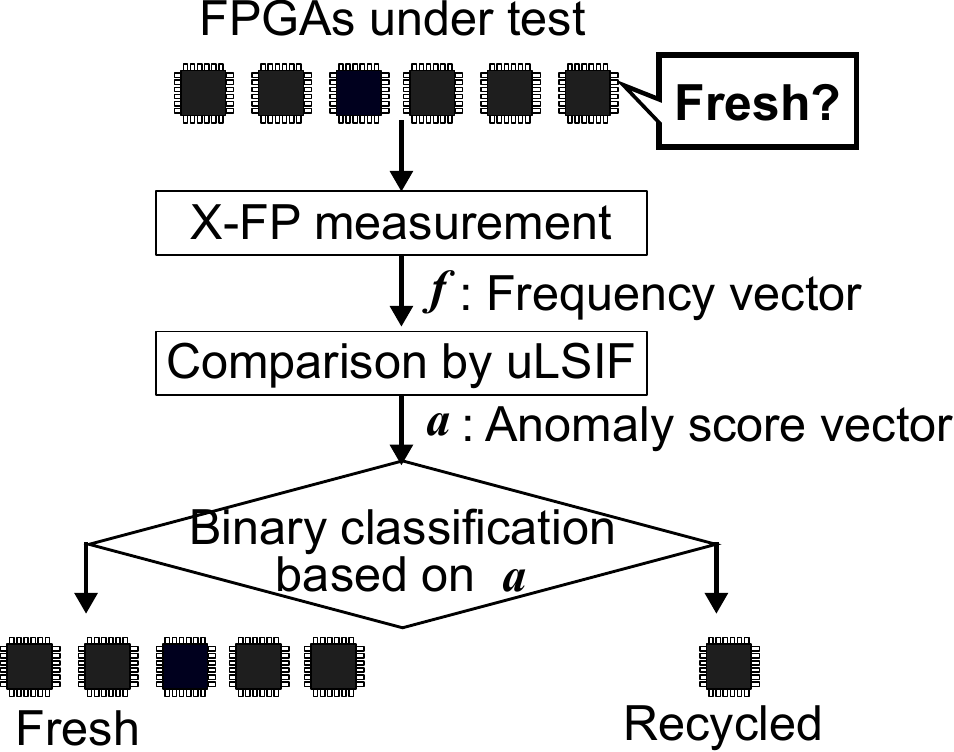}
    \caption{Overall flowchart of the proposed method.}
    \label{fig:flow}
\end{figure}

Our aim is to successfully identify recycled FPGAs without using a
\isakaFix{set of KFFs}. The proposed method involves
three steps: measurement, comparison based on uLSIF, and
classification, as shown in Fig.~\ref{fig:flow}. The frequencies of
target FPGAs are exhaustively characterized by
the X-FP measurement
and then \proof{based on the frequencies,} the anomaly score vectors are calculated \proof{using the} uLSIF.
Using the anomaly score vector, the FPGA is tested
\proof{and classified as either fresh or recycled.}

\subsubsection{Measurement}
\begin{figure}[t]
    \centering
    \includegraphics[width=0.9\linewidth]{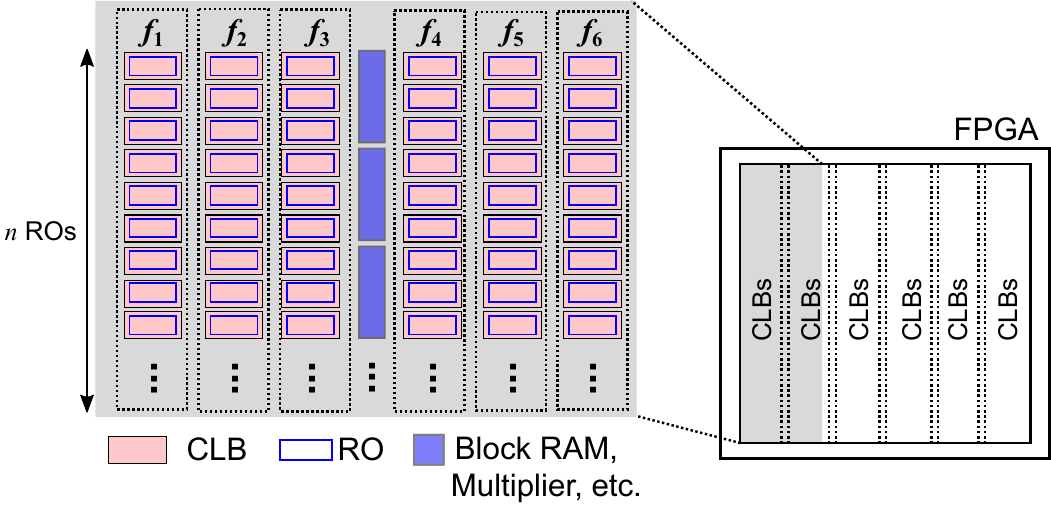}
    \caption{Overview of the RO array design. In the proposed method,
          each RO is designed using two CLBs, i.e., a CLB and its neighbor on the right.
          The frequency vector
          $\bm{f}$ is composed of the frequencies on the same column.}
    \label{fig:array}
\end{figure}

ROs are first designed and measured on
an FUT based on the exhaustive
path fingerprinting (X-FP) measurement method proposed
in~\cite{TCAD_Ahmed}. X-FP measurement is a combination 
of
two previous methods~\cite{ICCD2017_Jyothi,TVLSI2019_Alam}.
As opposed to XP measurement requiring a step of the target CLB
selection, X-FP measurement performs for all the LUTs paths of all the
CLBs in the FPGA, capturing the aging effect exhaustively.
An RO is designed by defining the
logic gates using a hardmacro function of a computer-aided design
(CAD) tool. As shown in Fig.~\ref{fig:array}, all the available CLBs
are analyzed through RO design.



\subsubsection{Comparison}
As mentioned above, ROs on adjacent CLBs
ideally have equal
frequency values. Recycled FPGAs violate this assumption \proof{by a} large difference.
Quantitative comparison of
these frequencies renders
it possible to detect recycled FPGAs without
requiring KFFs.  Note that
because the proposed method exhaustively analyzes all
the CLBs and LUTs \proof{as X-FP},
it is possible to analyze the effect of deterioration without any
omission by comparing all the columns.

In the proposed method, recycled FPGA detection is performed based on
the anomaly score calculated using
the uLSIF.
For calculating the anomaly score, the
RO frequencies
on the same column are defined as
$\bm{f}_i=\{f_{i;1},f_{i;2},\cdots,f_{i;n}\}$, where $f_{i;n}$ is the
RO frequency of the $n$-th RO in the $i$-th column, and \proof{the RO frequencies} are compared with the
adjacent column, as shown in Fig~\ref{fig:array}.
In this example,
$\bm{f}_1$ is compared with $\bm{f}_2$, $\bm{f}_2$ with $\bm{f}_3$, and
$\bm{f}_4$ with $\bm{f}_5$.
However,
modern FPGAs have several block RAMs (BRAMs) and
multiplier columns between the logic block areas; hence,
anomaly score calculation is not carried out across them (\proof{for example}, $\bm{f}_3$ with
$\bm{f}_4$) \proof{as} no similarity is guaranteed \proof{because of} the distance
between the columns.


The two frequency vectors of neighboring columns are
\proof{inputted} to the uLSIF as $\bm{f}$ and $\bm{f'}$ for calculating the anomaly scores. Given
all the comparison combinations,
the uLSIF returns the anomaly
scores for each calculation, for example, $M$ scores are obtained as
$\bm{A}=\{\bm{a}_1, \bm{a}_2, \cdots, \bm{a}_M\}$, when $M$ comparisons
are made.
It should be noted that
the comparisons perform in the same LUT paths configuration because
each path may have a different frequency trend.  Therefore, the uLSIF
procedure is iterated for all the LUT paths, \proof{in this case}, $2^{z-1}$ times per
one FPGA.



\subsubsection{Classification}
Finally, the FUT is classified as fresh or recycled based on the
anomaly scores. If the FUT was used previously, large scores should be
observed. However, it might be difficult to separate the scores for
short usage. In this case, a fresh/recycled threshold can be
determined if the information of a KFF set is available, but note that
the data volume is very little as with~\cite{TVLSI2019_Alam}.

\section{Experimental}\label{sec:exp}
To demonstrate the effectiveness of the proposed method, we conducted
experiments using the Xilinx Artix-7 FPGA \isa{(XC7A35T-ICPG236C)}~\cite{artix7}.
Based on the measurement results, the proposed recycled FPGA
detection method was compared to the conventional method
presented in~\cite{TVLSI2019_Alam}.
We used 35 FPGAs (FPGA-01 to FPGA-35) for this experiment. {\sntn{We
    here note that the 35 FPGAs were manufactured in different
    production lots because bought from different distributors at
    different timings.  Since the proposed method relies on the
    feature of the systematic component of the process variation, it
    is important to evaluate FPGAs with various process variations
    between lots.}}

\subsection{Setup}\label{sec:measure_setup}

\begin{figure}[!t]
    \centering \subfigure[s9234]{
        \includegraphics[width=0.28\linewidth]{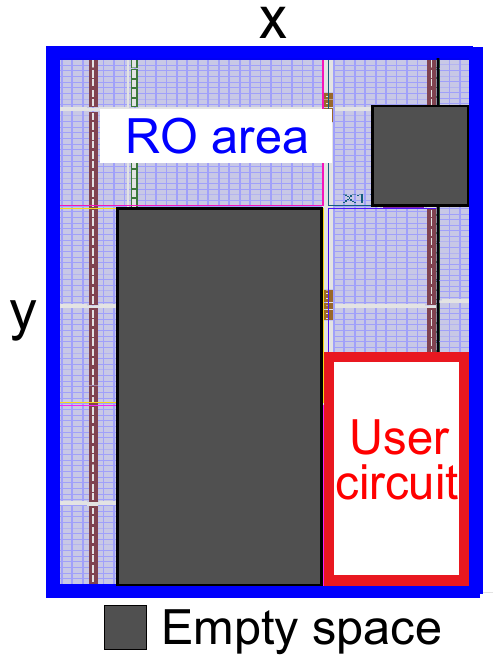}
        \label{fig:benchmark_s9234}
    }
    \hspace{3mm}
    \subfigure[RISC-V]{
        \includegraphics[width=0.28\linewidth]{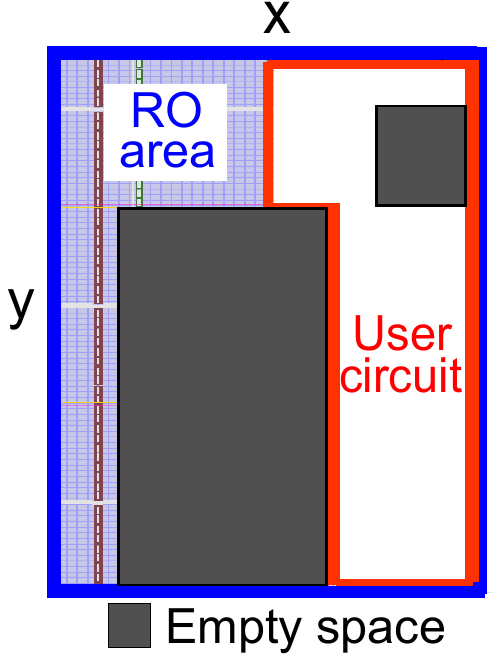}
        \label{fig:benchmark_riscv}
    }
    \caption{Floorplan of the Artix-7 FPGA with the user circuits.}
    \label{fig:benchmark}
\end{figure}

\begin{figure}[!t]
    \centering \subfigure[Path-2]{
        \includegraphics[width=0.32\linewidth]{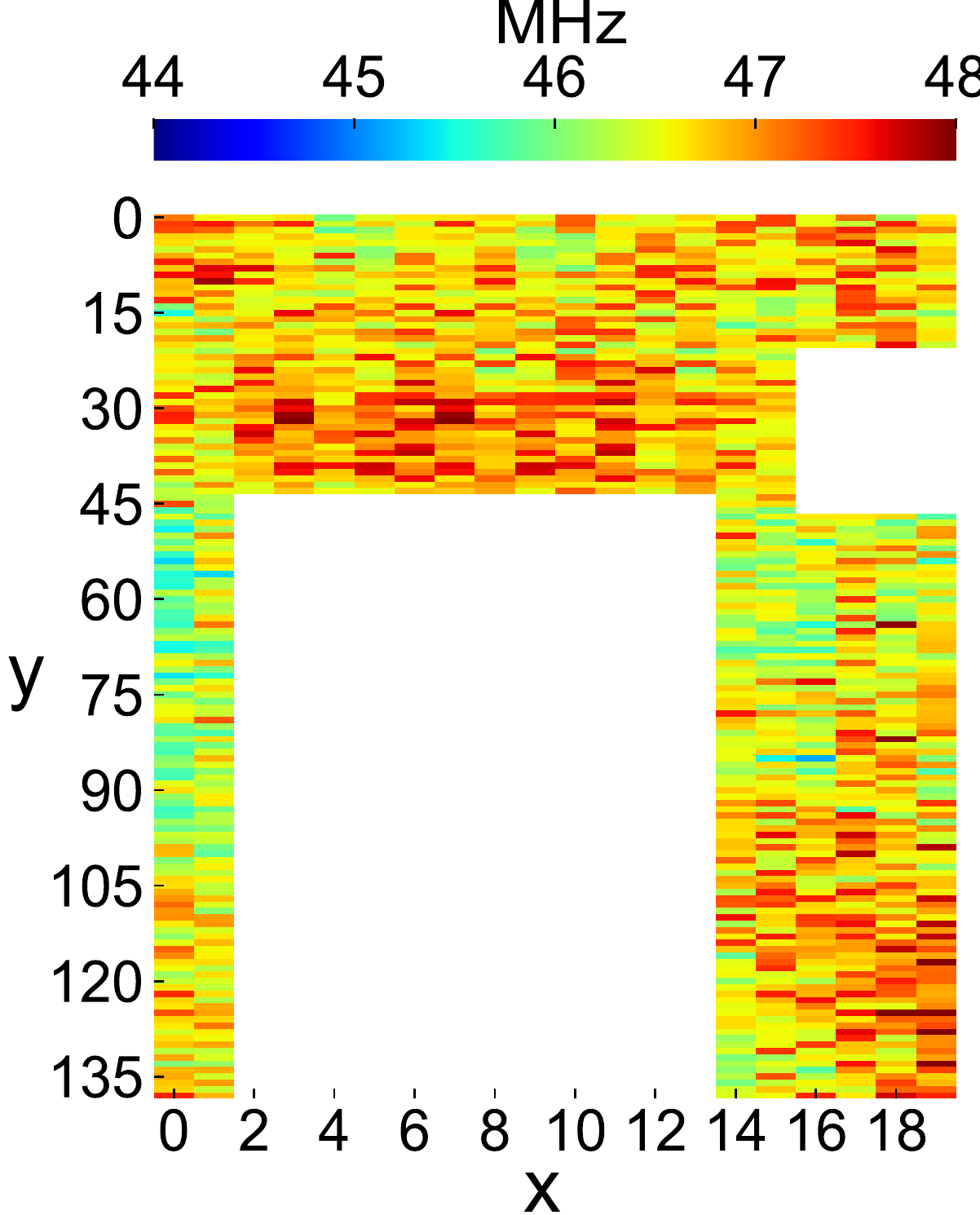}
        \label{fig:heatmap_fpga01path2}
    }
    \hspace{3mm} \subfigure[Path-4]{
        \includegraphics[width=0.32\linewidth]{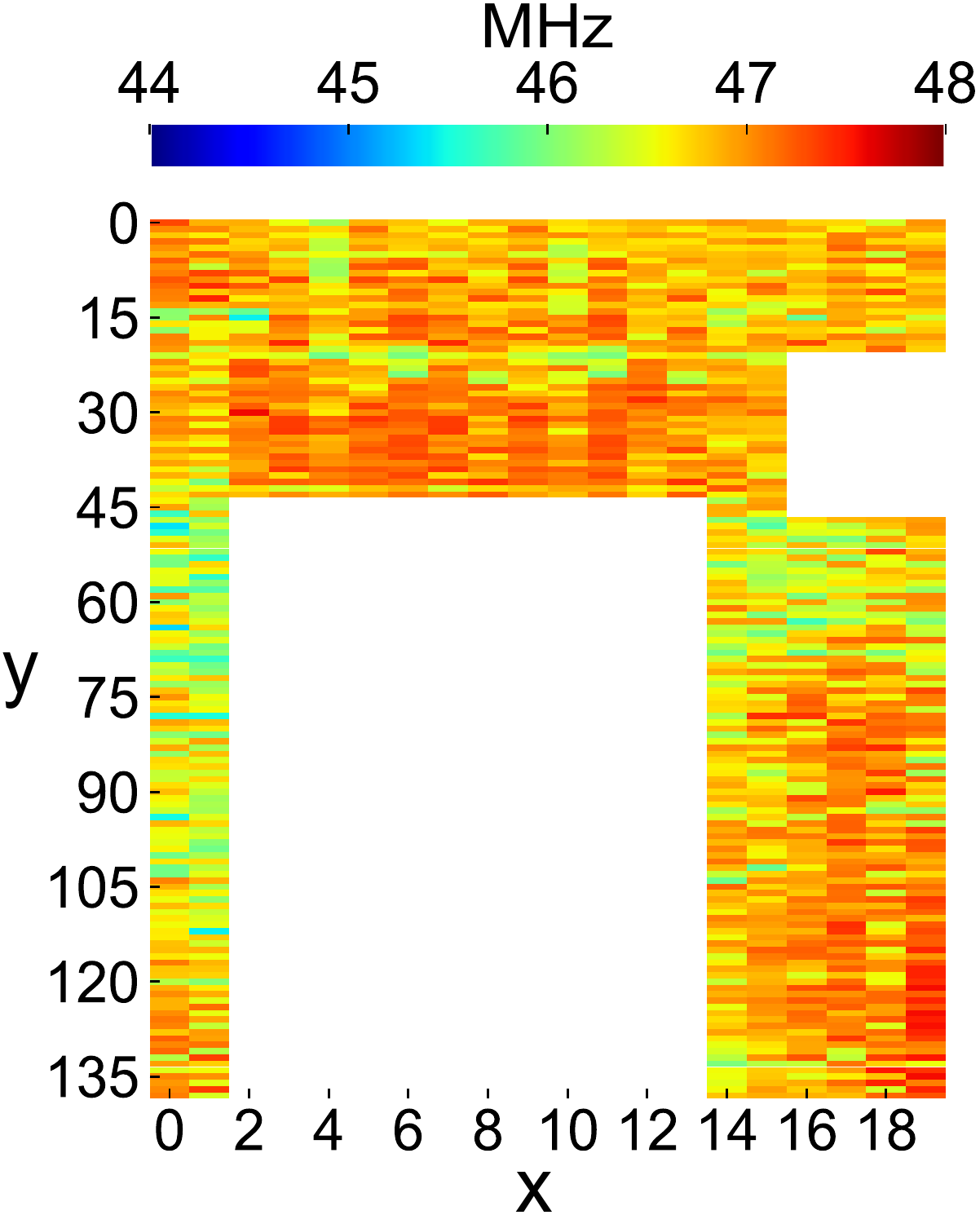}
        \label{fig:heatmap_fpga01path4}
    }
    \caption{\isakaFix{Frequency heatmaps of the fresh FPGA-01.}}
    \label{fig:heatmap_original1}
\end{figure}

\begin{figure}[!t]
    \centering \subfigure[Path-2]{
        \includegraphics[width=0.32\linewidth]{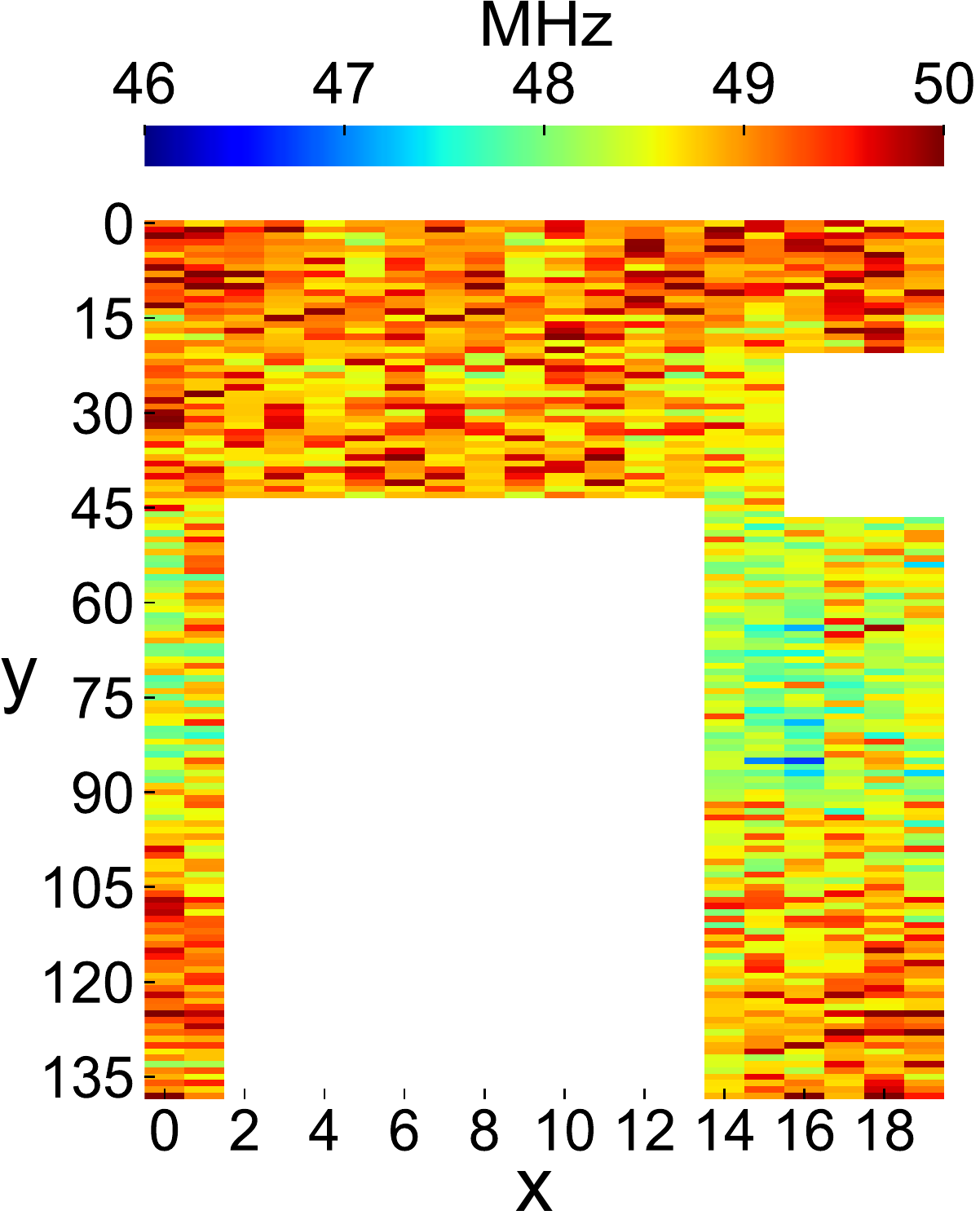}
        \label{fig:heatmap_fpag02path2}
    }
    \hspace{3mm} \subfigure[Path-4]{
        \includegraphics[width=0.32\linewidth]{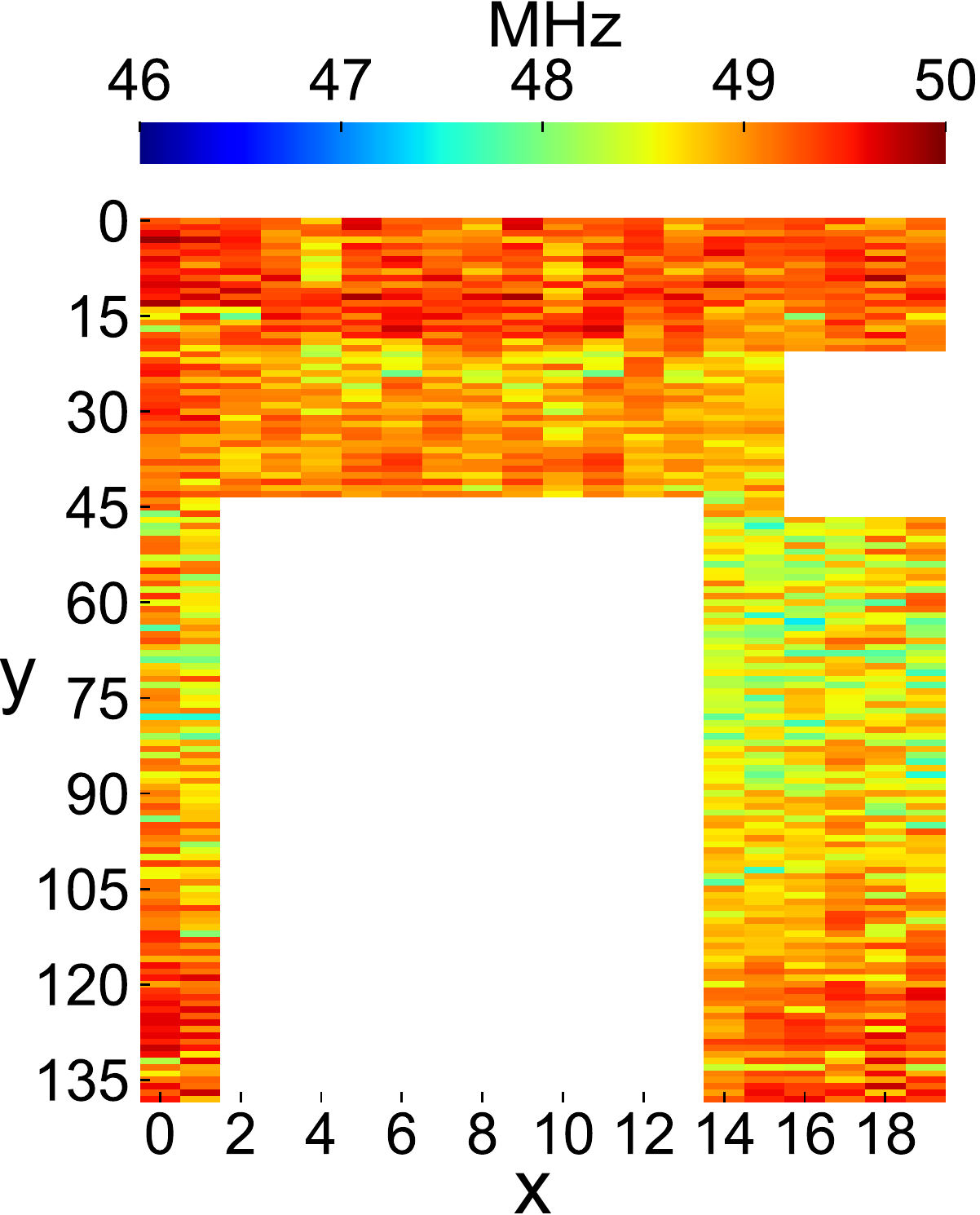}
        \label{fig:heatmap_fpag02path4}
    }
    \caption{\isakaFix{Frequency heatmaps of the fresh FPGA-02.}}
    \label{fig:heatmap_original2}
\end{figure}

In the floorplan of the Artix-7 depicted in Fig.~\ref{fig:benchmark},
most of the area is occupied by CLBs, and there is some
empty space area \proof{with} no logic resource.
In addition, \isakaNew{columns of BRAMs can be found.}
\isakaNew{An on-chip measurement system with 15-stage of ROs (two CLBs) was implemented using eight
LUTs on the Artix-7 CLB.}
The RO was designed using XNOR or XOR logic gates.
\isakaNew{An Artix-7 FPGA has 6-input LUTs, and in the configuration of RO
using XNOR or XOR\proof{,} which make a total of 32 $(=2^{6-1}$) LUT
paths (path-0 to path-31) can be measured for each CLB.}
A total of 42,112 ROs were placed on a geometric grid,
and their frequencies were measured using a counter circuit\footnote{The verilog code for the X-FP measurement is available at \url{https://github.com/yuya-isaka/X-FP}.}.
By \proof{retaining} the same internal routing through the hardware macro
modeling function of Vivado (Xilinx CAD tool)~\cite{vivado}, the logic
resources and structure for each RO were placed in two CLBs
such that the frequency variation caused by
the difference in internal routing can be minimized. In the layout, 10
comparisons were made ($M=10$) because \proof{of} a couple of BRAM
columns. The heatmaps of the measured frequencies of path-2 and
path-4 for the fresh FPGA-01 and FPGA-02 are shown in
Figs.~\ref{fig:heatmap_original1} and~\ref{fig:heatmap_original2}. It
can be observed that the frequency distributions in the fresh and aged
fingerprints have a smooth variation along \isaka{with} the coordinates and
\proof{are differing frequencies.}

\begin{table}[!t]
    \centering
    \caption{List of the aged FPGAs}
    \label{tab:aged_list}
    \begin{tabular}{c|c|c}
      \hline
      Device ID & User circuit  & Aging time, $t$  \\ \hline
      FPGA-01    & s9234  & 6 \\ 
      FPGA-02    & s9234  & 6 \\ 
      FPGA-03    & s9234  & 3 \\ 
      FPGA-04    & s9234  & 2 \\ 
      FPGA-05    & s9234  & 1 \\ 
      FPGA-06    & RISC-V  & 6 \\
      FPGA-07    & RISC-V  & 3 \\
      FPGA-08    & RISC-V  & 2 \\
      FPGA-09    & RISC-V  & 1 \\ \hline
    \end{tabular}
\end{table}

\begin{figure}[!t]
    \begin{center}
        \includegraphics[width=.57\columnwidth]{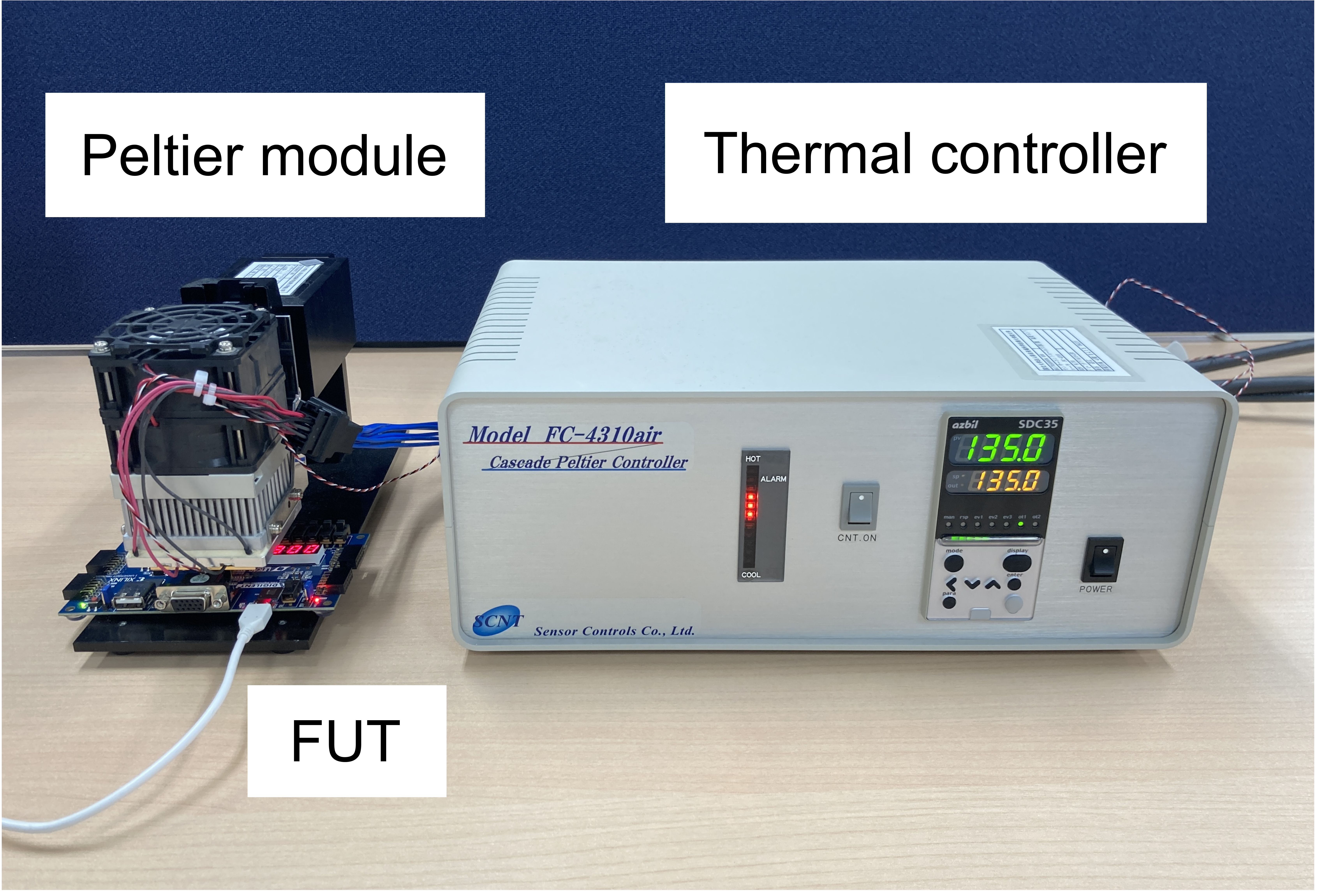}
    \end{center}
    \caption{Experimental setup for accelerating the aging process.}
    \label{fig:equip}
\end{figure}

\begin{figure}[!t]
    \centering
    \includegraphics[clip, width=0.73\linewidth]{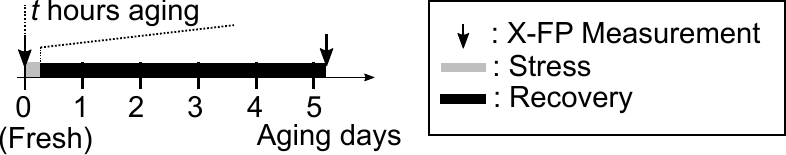}
    \caption{Aging schedule and measurement timing.
      \proof{We} changed the aging time $t$ to evaluate under
      various aging scenarios.}
    \label{fig:schedule}
\end{figure}

In the experiments, \isakaNew{nine} FPGAs (FPGA-01 \proof{to} FPGA-09)
were aged \proof{and used} as the recycled FPGAs among the 35 FPGAs
employed for testing as listed in Table~\ref{tab:aged_list}.  The
s9234 benchmark circuit~\cite{ISCAS1989_Brglez} and a RISC-V processor
run as a user circuit on {\shintani{the \isakaNew{nine} FPGAs for
    FPGA-01 to FPGA-05 and FPGA-06 to \isakaNew{FPGA-09}}},
respectively. While running the circuits, the FPGAs were heated up to
\proof{135\,$^\circ$C using} a Peltier module with a thermal controller, as
shown in Fig.~\ref{fig:equip}, to accelerate the aging process.  The
aging schedule includes stress and recovery phases, and we \proof{changed} the
stress time $t$ to evaluate the proposed method under various aging
scenarios as shown in Fig.~\ref{fig:schedule}.

\isakaNew{FPGA-01 and FPGA-02 were aged for 6\,\proof{h} ($t=6$). FPGA-03, FPGA-04,
and FPGA-05 were aged for 3, 2, and 1\,\proof{h} ($t=3,2,1$), respectively.}
\isakaNew{The s9234 circuit continuously applies a pseudo-random
number pattern from a 16-bit linear feedback shift register (LFSR) at 100\,MHz.}
\isakaNew{The user circuit, including the LFSR and the frequency counter, was placed
at the bottom right corner on the FPGA as shown in Fig.~\ref{fig:benchmark_s9234}.}
FPGA-06 to FPGA-09 with the RISC-V processor were aged for
  $t=6,3,2,1$, respectively.
{\shintani{The Dhrystone program, \proof{widely used to evaluate}
    performance, repeatedly runs on the RISC-V
    processor at \isakaNew{100}\,MHz during the aging.}}  We used a 32-bit instruction set (RS32IM) and developed the
program using a GNU Compiler Collection (GCC) \proof{tool chain} provided in~\cite{gcc}\footnote{The RISC-V environment is available at \url{https://github.com/yuya-isaka/RV32}.}.
\isakaNew{The user circuitry, including the RISC-V processor and the frequency counter,
was placed on the right side on the FPGA, as shown in Fig.~\ref{fig:benchmark_riscv}.}
In the measurement schedule, measurements were made after a
five-day recovery phase.

\isakaNew{The actual \proof{FPGA running time} relative to the degradation time
due to thermal acceleration can be estimated from the usage conditions
of the thermal device~\cite{Reconfig2015_Gehrer}.
The thermal acceleration factor $F_{\text{T}}$ can be calculated
\proof{as~\cite{maes2012experimental}:}}
\begin{equation}
    F_{\text{T}}=\mathrm{e}^{\frac{E_{\rm a}}{k}(\frac{1}{T_{\text{op}}}-\frac{1}{T_{\text{stress}}})},
    \label{eq:thermal}
\end{equation}
\isakaNew{where, the activation energy $E_{\rm a}=0.5$\,eV, Boltzmann
  constant $k=8.62\times10^{-5}$\,eV/K, nominal operating temperature
  $T_{\text{op}}=313$\,K (40\,$^\circ$C), and stress temperature
  $T_{\text{stress}}=408$\,K (135\,$^\circ$C), respectively.}
\isakaNew{Note that it is difficult to \proof{accurately} estimate the actual operating
  \proof{time since our experiment's degradation schedule
  includes a recovery phase.}  We can roughly estimate
  the accelerated aging of 6, 3, 2, and 1\,\proof{h} as the actual
  operating time of 18, 9, 6, and 3\,\proof{d}, respectively.}

Note that a similar aging acceleration \isaka{was} conducted for 24\,\proof{h}
in~\cite{TCAD_Ahmed}.  \proof{With their study}, it \proof{is} difficult to
\proof{accurately} detect \proof{the} aged FPGAs\proof{,} even by the previous method based
on a set of KFFs.  \proof{Although a} \isakaNew{similar} aging
acceleration \proof{method} using a temperature controller \proof{was performed}
in~\cite{TVLSI2019_Alam}, the degradation \proof{in their method} was greater because
\proof{they did not consider a recovery phase.} In our experiment, a recovery phase
of \proof{over} four days was \proof{included to reflect} a realistic scenario
\proof{in the recycling of FPGAs}. In addition, the aging time was longer
than that of our \proof{method}. \proof{Therefore}, the recycled FPGA detection \proof{method}
\proof{undertaken} in \proof{our study} is \proof{a greater challenge than the other previous works~\cite{TVLSI2019_Alam,TCAD_Ahmed}}.

We implemented the uLSIF and $k$-means++ clustering
in the Python language.

\subsection{Results}
\subsubsection{Measurement results}

\begin{figure}[!t]
    \centering \subfigure[\isakaNew{\proof{6-h} aged FPGAs (FPGA-01 and FPGA-02)}]{
        \includegraphics[width=0.48\linewidth]{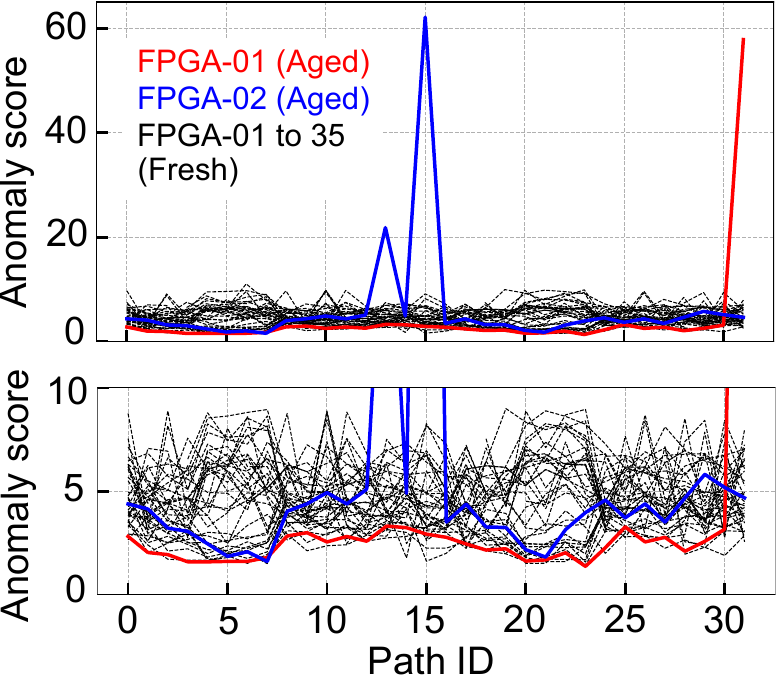}
        \label{fig:6aged_s9234}
    }
    \hspace{-4mm}
    \subfigure[\isakaNew{\proof{3-h} aged FPGA (FPGA-03)}]{
        \includegraphics[width=0.48\linewidth]{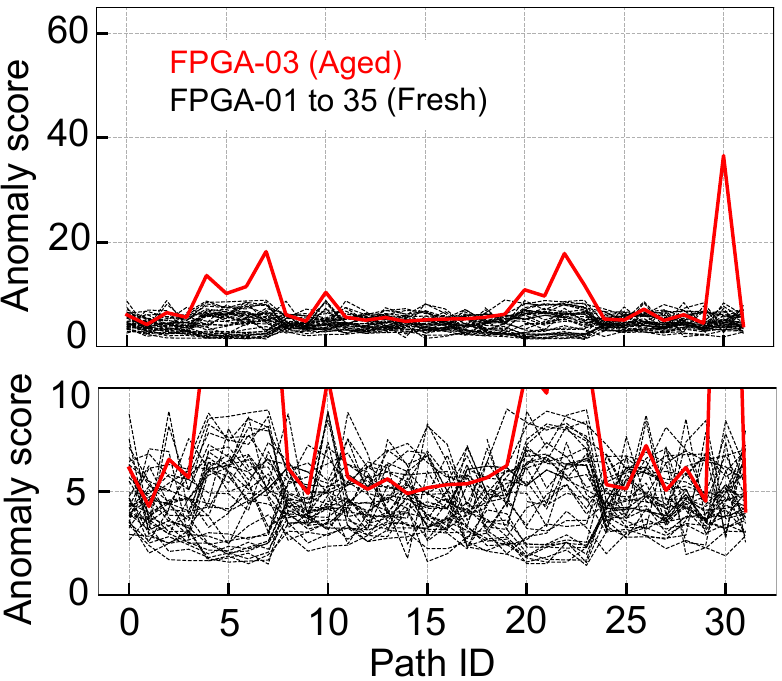}
        \label{fig:3aged_s9234}
    }
    \subfigure[\isakaNew{\proof{2-h} aged FPGA (FPGA-04)}]{
        \includegraphics[width=0.48\linewidth]{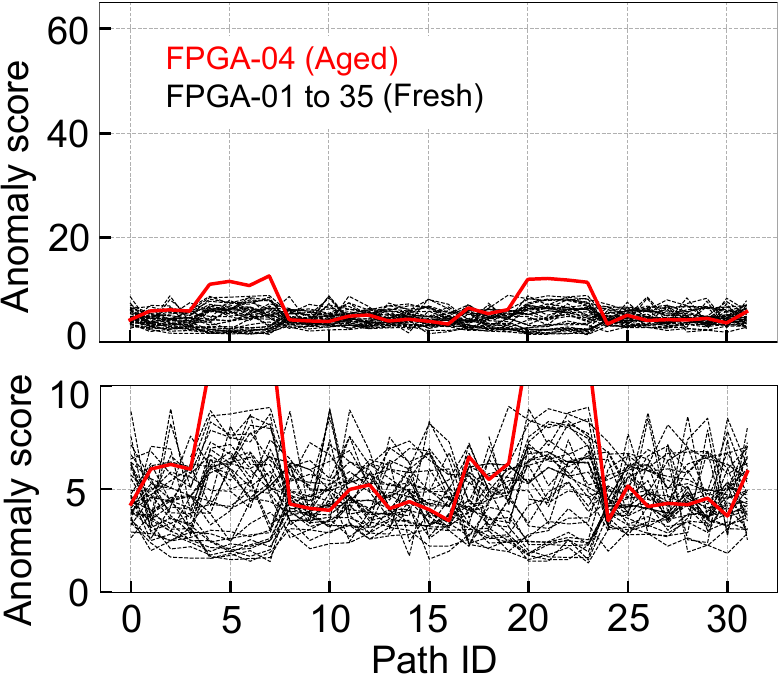}
        \label{fig:2aged_s9234}
    }
    \hspace{-4mm}
    \subfigure[\isakaNew{\proof{1-h} aged FPGA (FPGA-05)}]{
        \includegraphics[width=0.48\linewidth]{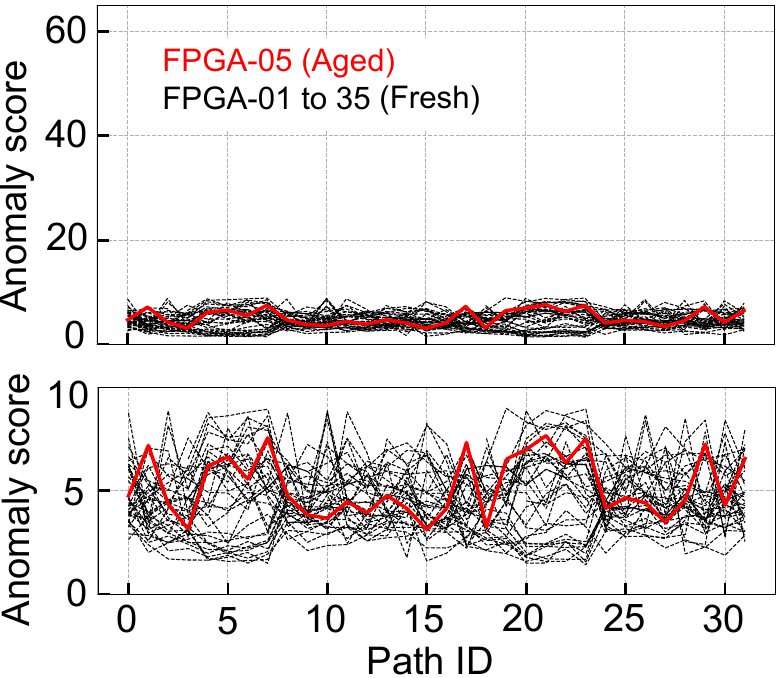}
        \label{fig:1aged_s9234}
    }
    \caption{{\shintani{Maximum anomaly scores of the 32 LUT paths for
        the 35 fresh FPGAs and five aged FPGAs (FPGA-01 to FPGA-05) with s9234.}}}
    \label{fig:anomaly_s9234}
\end{figure}

\begin{figure}[!t]
  \centering
  \subfigure[\isakaNew{\proof{6-h} aged FPGAs (FPGA-06)}]{
        \includegraphics[width=0.47\linewidth]{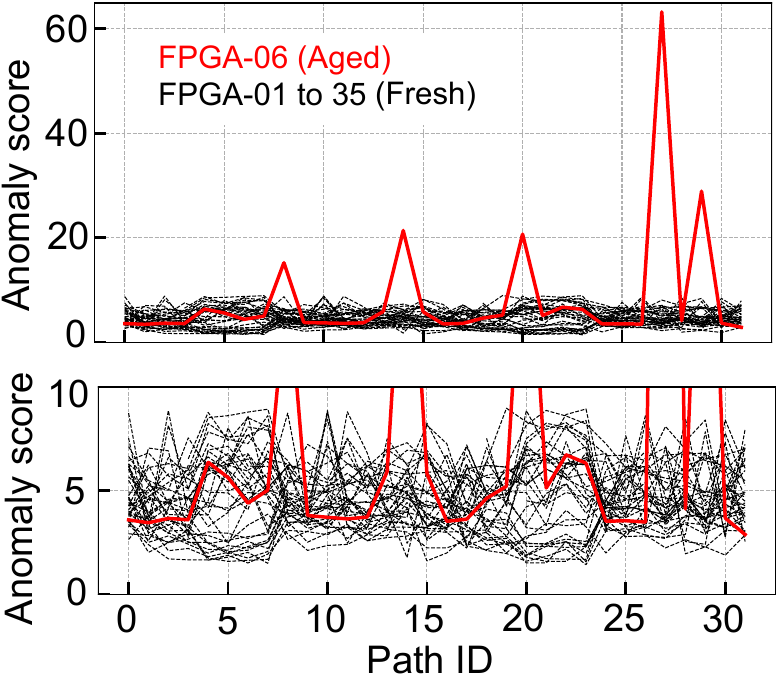}
        \label{fig:6aged_riscv}
    }
    \hspace{-4mm}
    \subfigure[\isakaNew{\proof{3-h} aged FPGA (FPGA-07)}]{
        \includegraphics[width=0.47\linewidth]{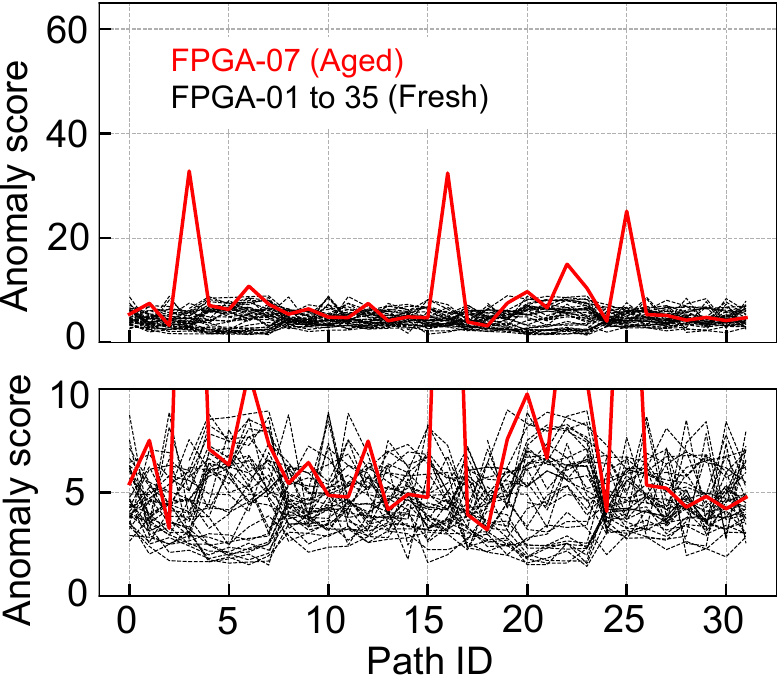}
        \label{fig:3aged_riscv}
    }
    \subfigure[\isakaNew{\proof{2-h} aged FPGA (FPGA-09)}]{
        \includegraphics[width=0.47\linewidth]{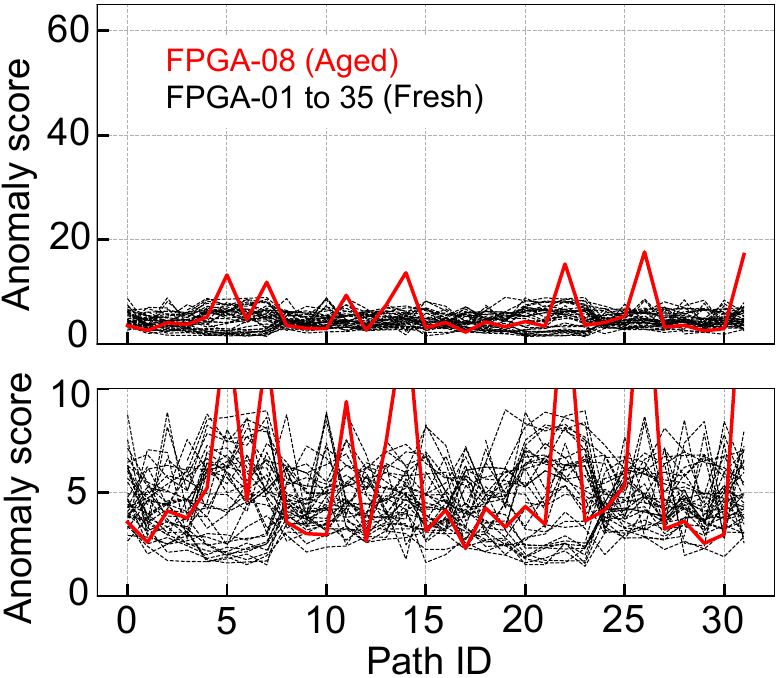}
        \label{fig:2aged_riscv}
    }
    \hspace{-4mm}
    \subfigure[\isakaNew{\proof{1-h} aged FPGA (FPGA-08)}]{
        \includegraphics[width=0.47\linewidth]{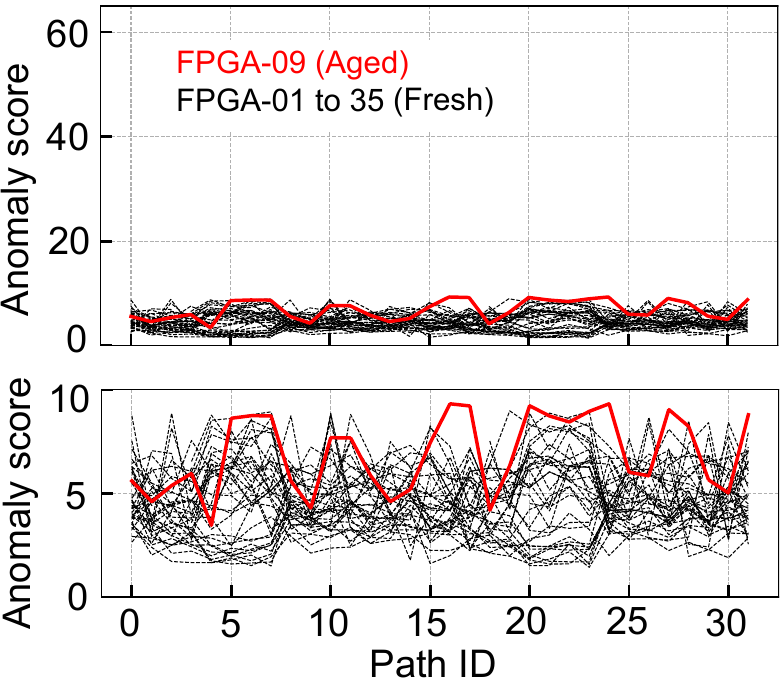}
        \label{fig:1aged_riscv}
    }
    \caption{{\shintani{Maximum anomaly scores of the 32 LUT paths for
        the 35 fresh FPGAs and four aged FPGAs (FPGA-06 to FPGA-09) with the RISC-V processor.}}}
    \label{fig:anomaly_riscv}
\end{figure}

First, we \proof{presented} the anomaly scores obtained \proof{from our} proposed method.
Fig.~\ref{fig:anomaly_s9234} displays the results of the five aged FPGAs of
s9234 (FPGA-01 to FPGA-05) and fresh FPGAs (FPGA-01 to FPGA-35).
The horizontal axis \proof{\isaka{represents} path IDs;} from path-0 to path-31.
The maximum anomaly score of the 10 comparisons (\proof{that is},
$\max_M(\bm{A})$ for each path) for the 32 LUT paths is \proof{also represented}.
For example, in Fig.~\ref{fig:6aged_s9234}, while the anomaly scores
for the fresh FPGAs \proof{ranged} from 1.0 to 10.0, several remarkably high
anomaly scores (\proof{such as}, above 20) were observed for the \proof{6-h} aged
FPGAs. \proof{The increase} from 1.0 to 10.0
comes from the random component in the process variation. Here, we
\proof{noted} that the shorter the aging time, the lower the
anomaly scores.
\isakaNew{Since high anomaly scores are observed in
  Figs.~\ref{fig:6aged_s9234},~\ref{fig:3aged_s9234}, and~\ref{fig:2aged_s9234},
  it is expected that
  the aged FPGAs \proof{are accurately} detected.} \proof{However,}
  it is not \proof{easy} to detect the \proof{1-h} aged shown in
Fig.~\ref{fig:1aged_s9234} \proof{because} the anomaly scores of recycled and fresh
\proof{one are almost the same}.

Similarly, Fig.~\ref{fig:anomaly_riscv} shows the results of the
four aged FPGAs of the RISC-V processor (FPGA-06 to FPGA-09) and fresh
FPGAs.
\proof{Even with} these \proof{results, the} obvious abnormal scores were
observed for the \proof{6-h}, \proof{3-h}, and \proof{2-h} aged FPGAs, the anomaly
score of the \proof{1-h} aged FPGA \proof{was almost equal to} the fresh ones
and we \proof{found it} difficult to identify the difference caused by
the aging.


\begin{figure}[!t]
  \centering
  \subfigure[Fresh FPGA-01]{
    \includegraphics[width=0.3\linewidth]{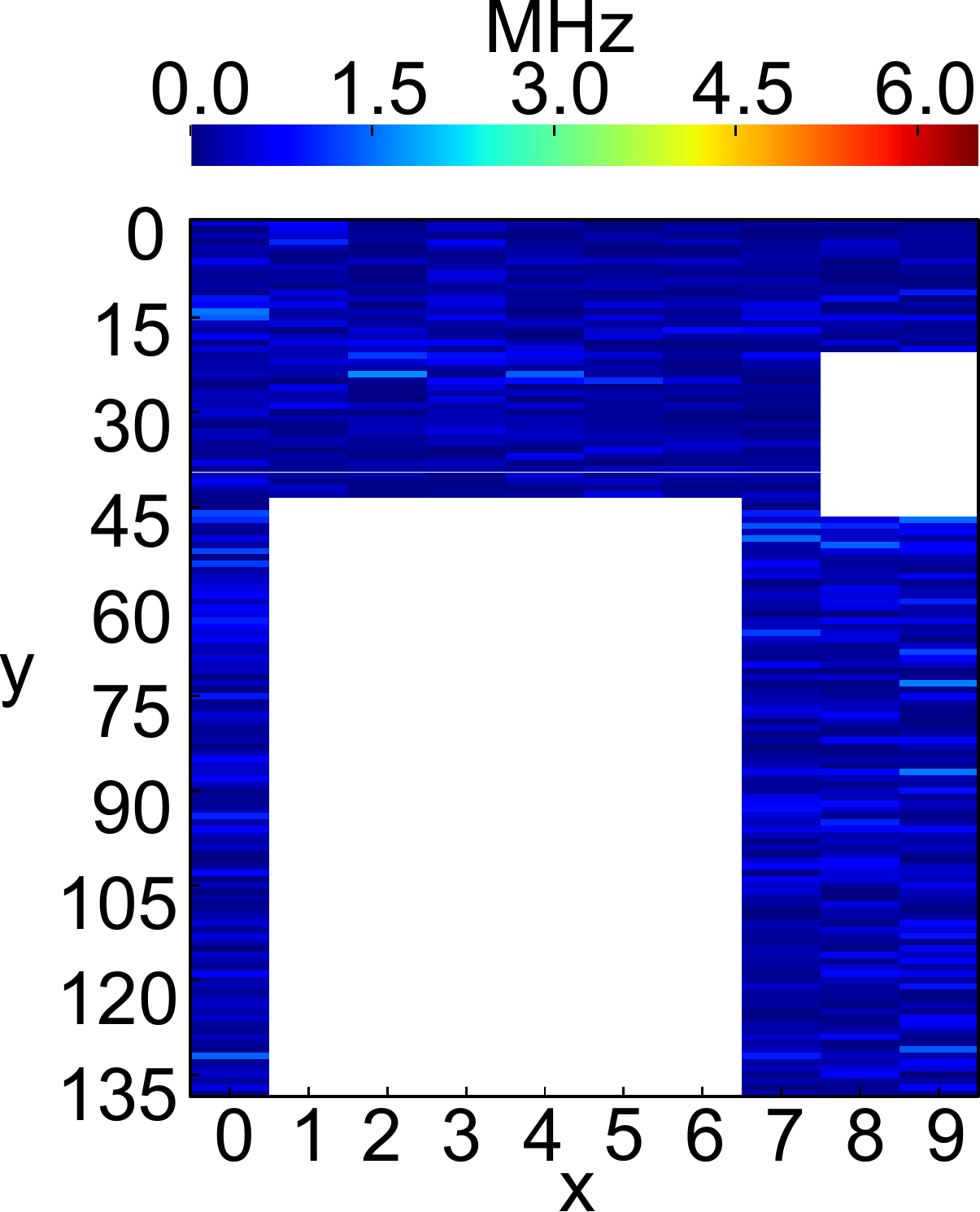}
    \label{fig:fresh01}
  }
  \subfigure[Fresh FPGA-02]{
    \includegraphics[width=0.3\linewidth]{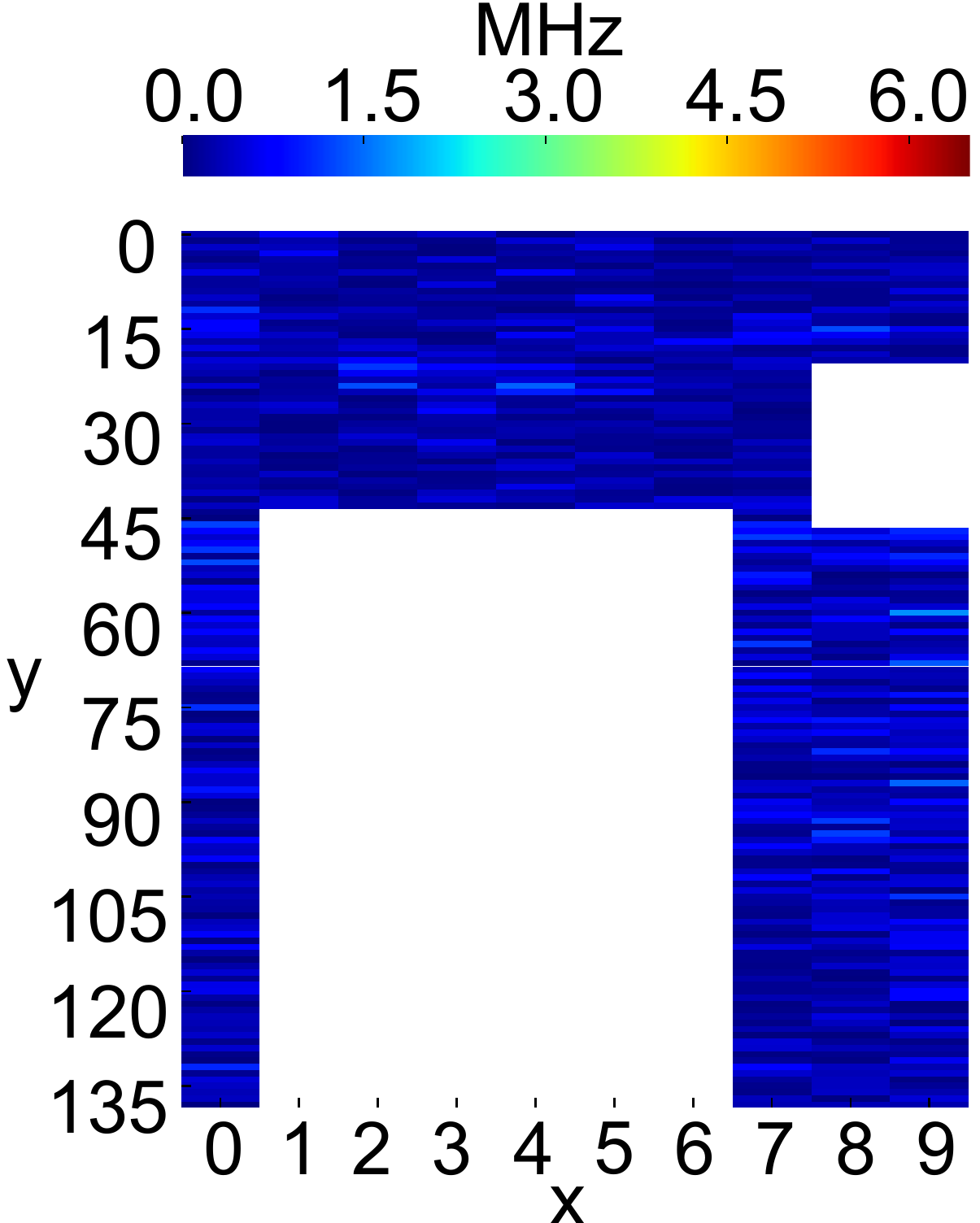}
    \label{fig:fresh02}
  }
  \caption{Frequency heatmaps of the residual frequencies of the column comparison for the fresh FPGA-01 and FPGA-02.}
  \label{fig:fresh_residual}
\end{figure}

Fig.~\ref{fig:fresh_residual} depicts the frequency heatmaps of FPGA-01 and
FPGA-02 when fresh, respectively. In the heatmaps, the residuals of
the two columns are shown using the coordinates of the RO on the
left-hand side in the compared column. \proof{The} paths with the
highest scores in Fig.~\ref{fig:anomaly_s9234} are shown.  As
expected, \proof{because} the frequencies in the neighboring columns are
similar, those residuals are almost zero. Although \proof{some} of the
fresh FPGAs, \proof{such as} FPGA-03 and FPGA-04, are \proof{not included}, similar results
can be \proof{obtained}.

\begin{figure}[!t]
  \centering
  \subfigure[Aged FPGA-01]{
    \includegraphics[width=0.3\linewidth]{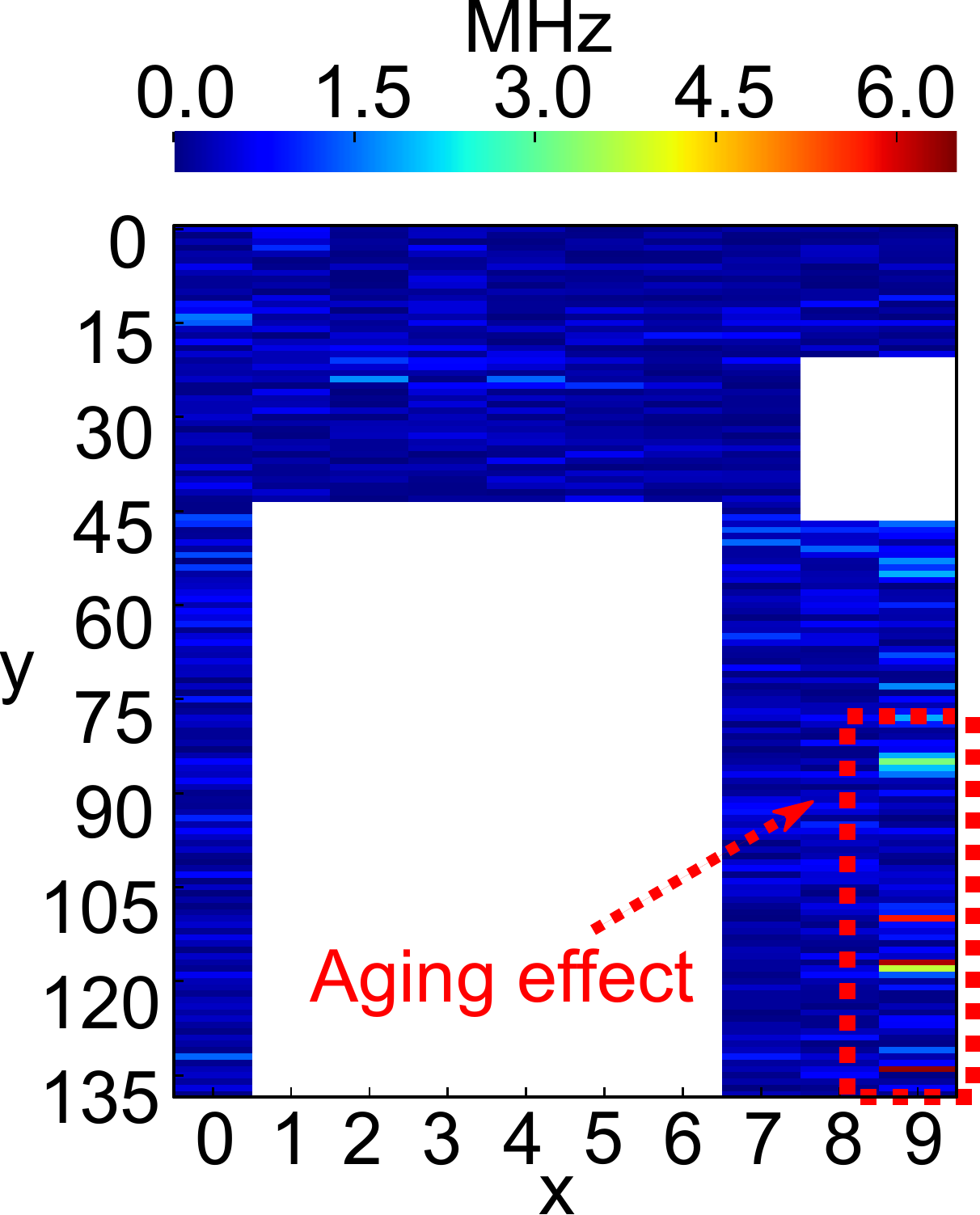}
    \label{fig:aged01}
  }
  \hspace{-3mm}
  \subfigure[Aged FPGA-02]{
    \includegraphics[width=0.3\linewidth]{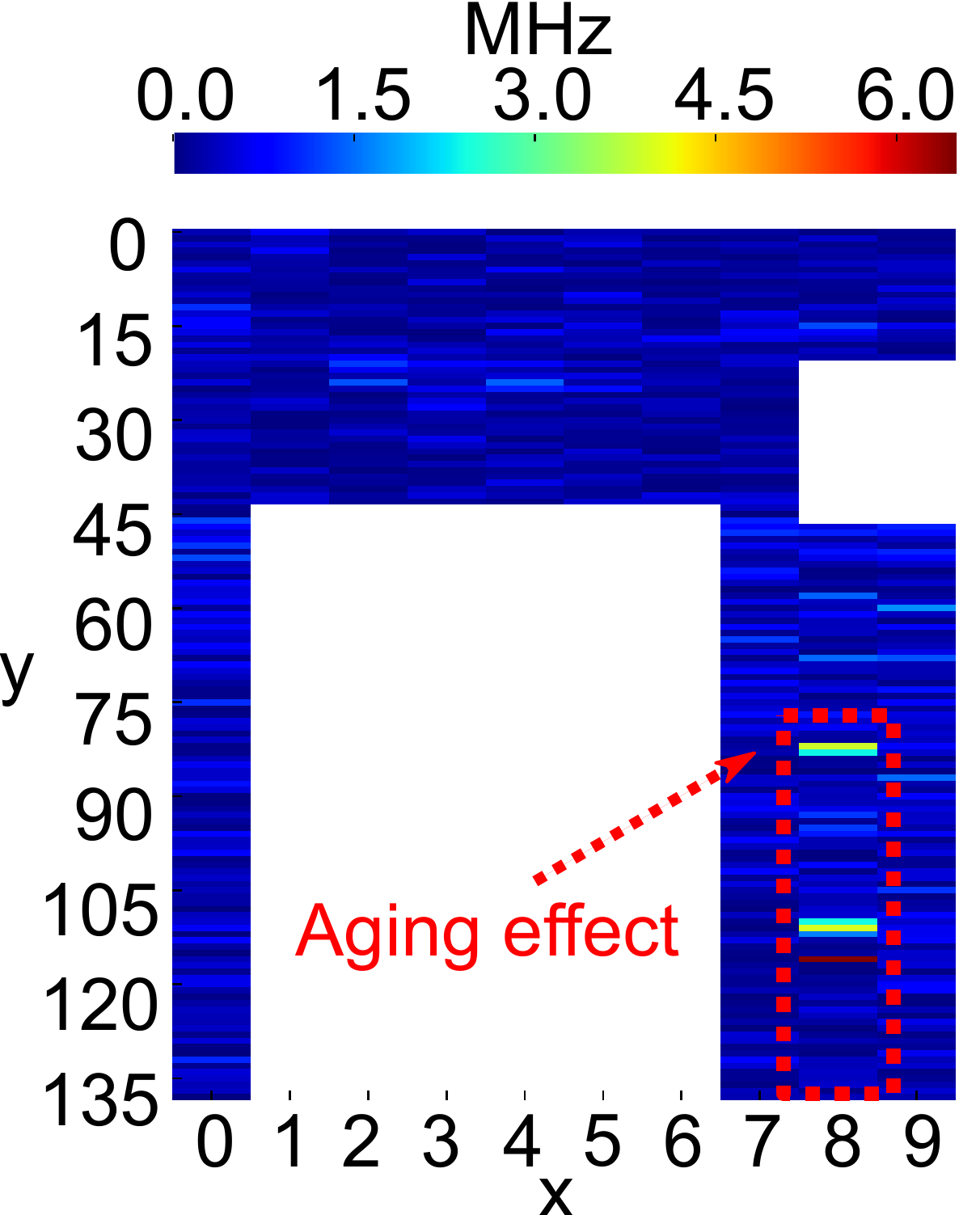}
    \label{fig:aged02}
  }
  \hspace{-3mm}
  \subfigure[Aged FPGA-03]{
    \includegraphics[width=0.3\linewidth]{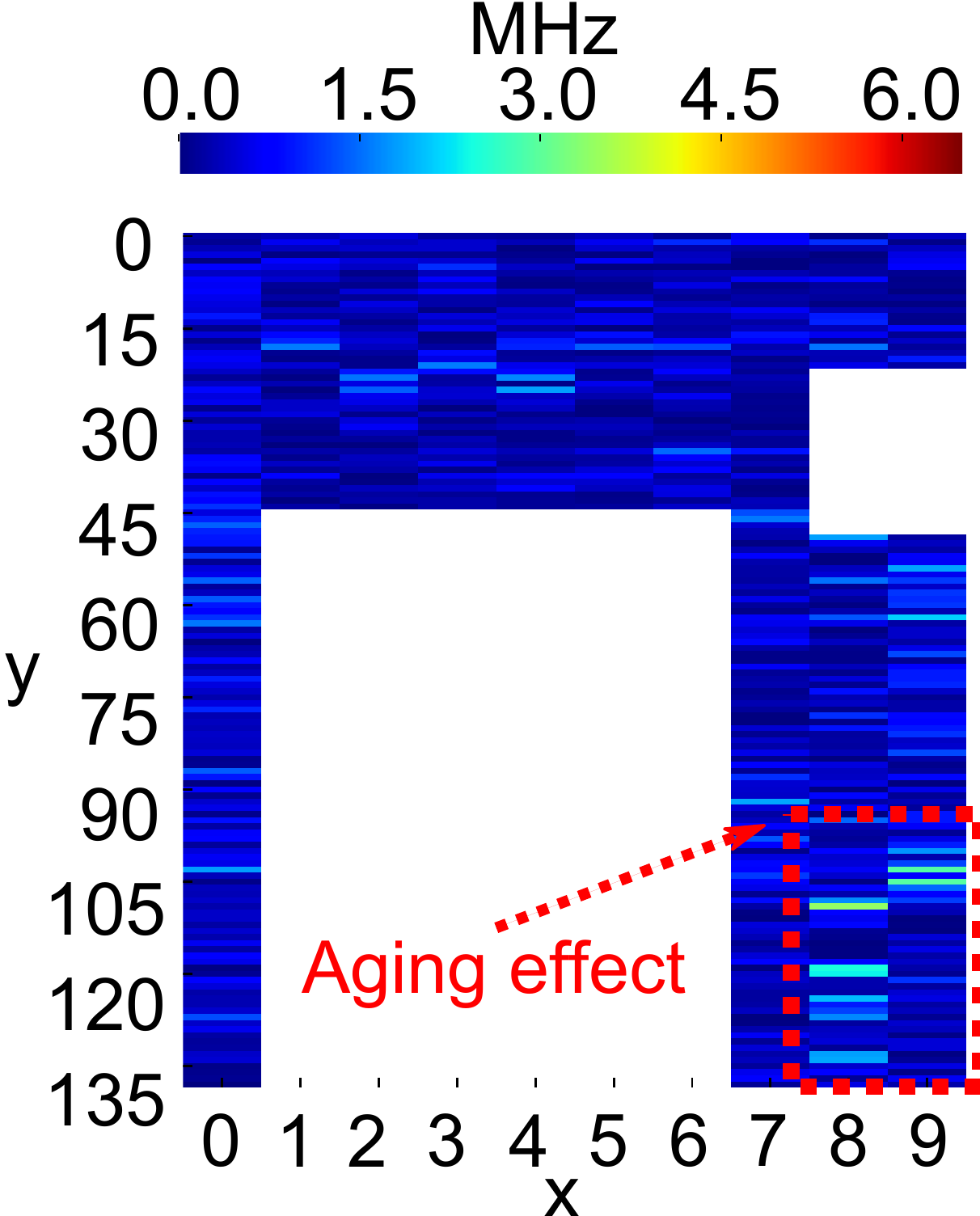}
    \label{fig:aged03}
  }
  \hspace{-3mm}
  \subfigure[Aged FPGA-04]{
    \includegraphics[width=0.3\linewidth]{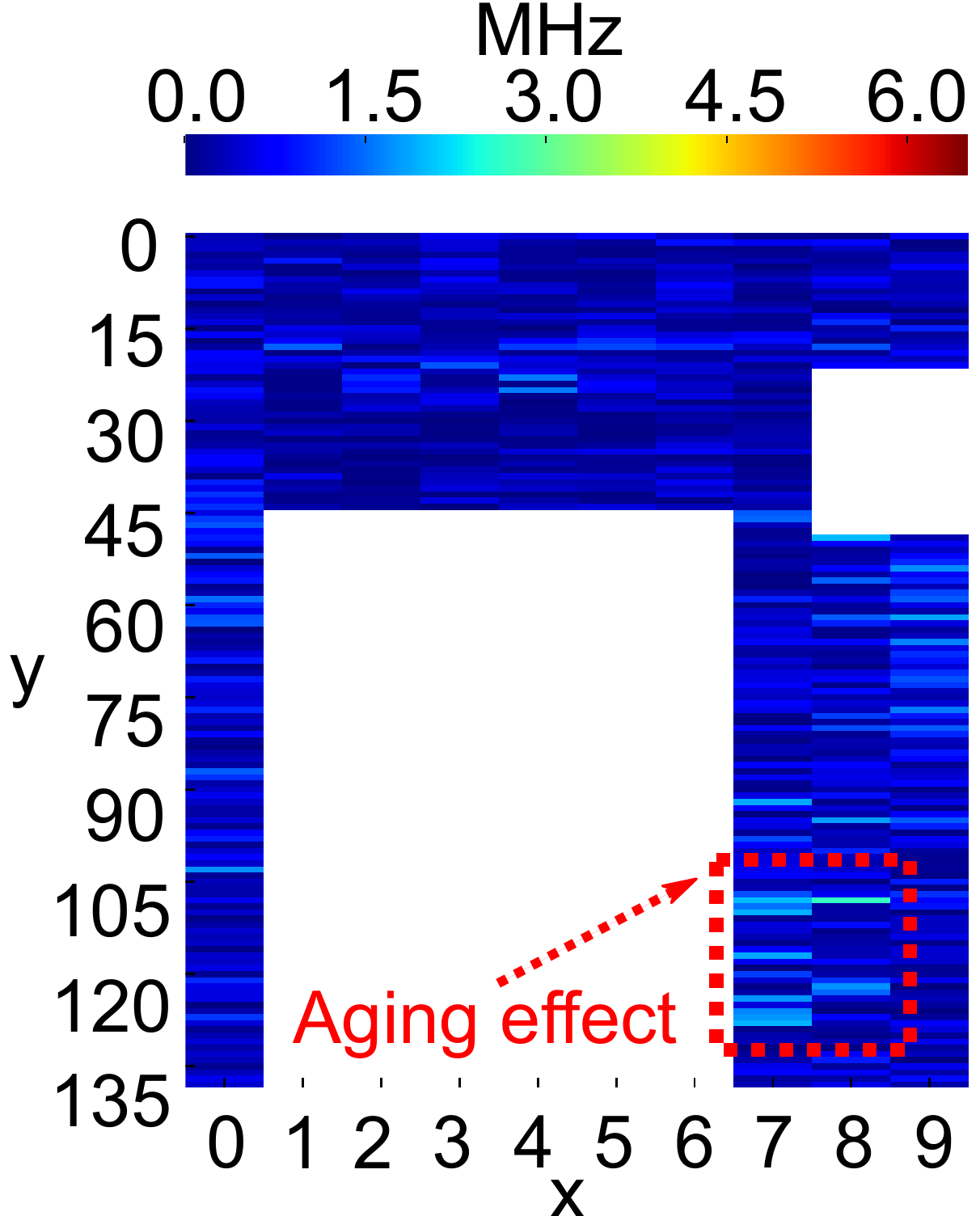}
    \label{fig:aged04}
  }
  \hspace{-3mm}
  \subfigure[Aged FPGA-05]{
    \includegraphics[width=0.3\linewidth]{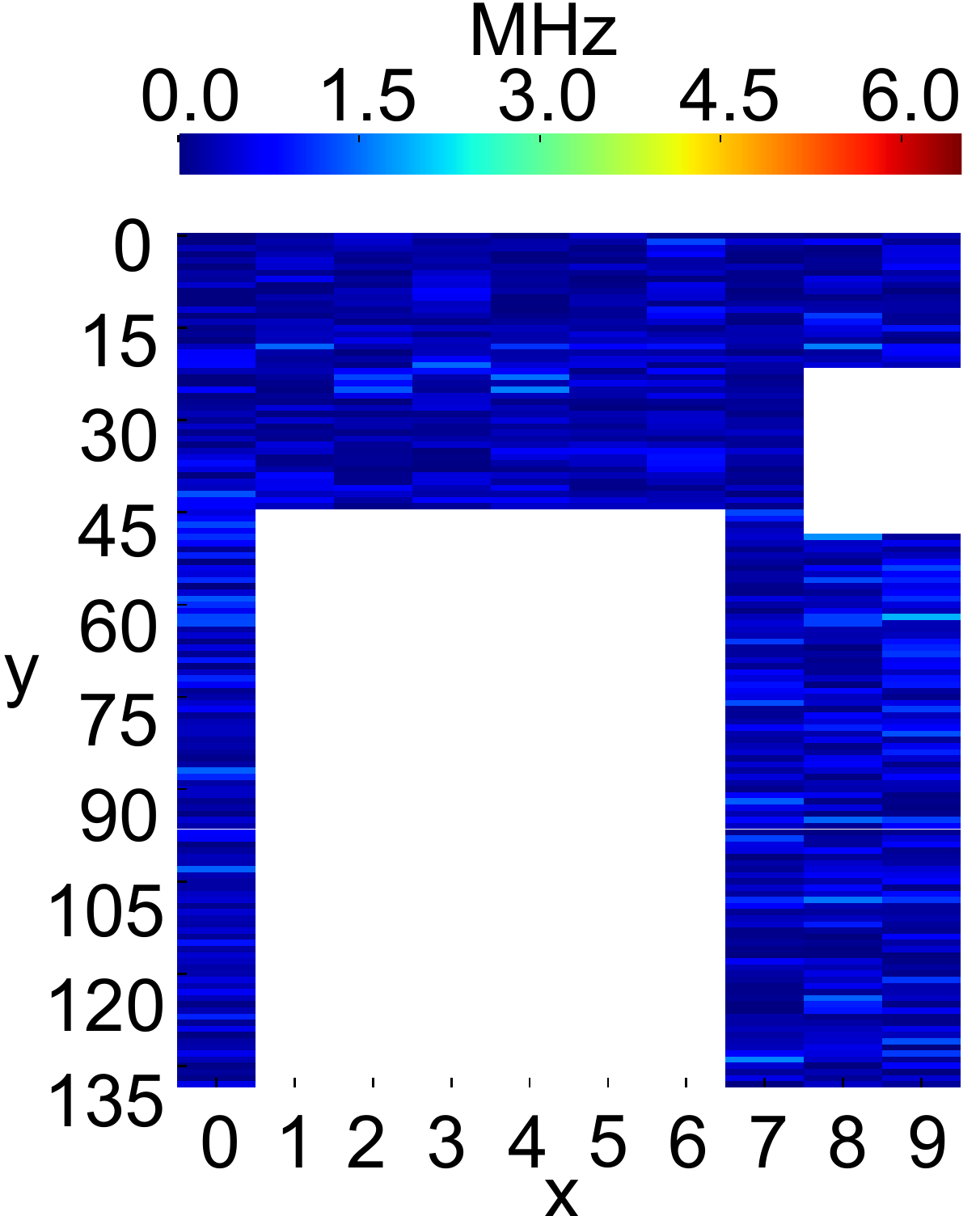}
    \label{fig:aged05}
  }
  \caption{\isakaNew{Frequency heatmaps of the residual frequencies of the column comparison for the aged FPGA-01 to FPGA-05.}}
  \label{fig:aged_residual}
\end{figure}

\begin{figure}[!t]
  \centering
  \subfigure[\proof{6-h} aged FPGA-06)]{
    \includegraphics[width=0.3\linewidth]{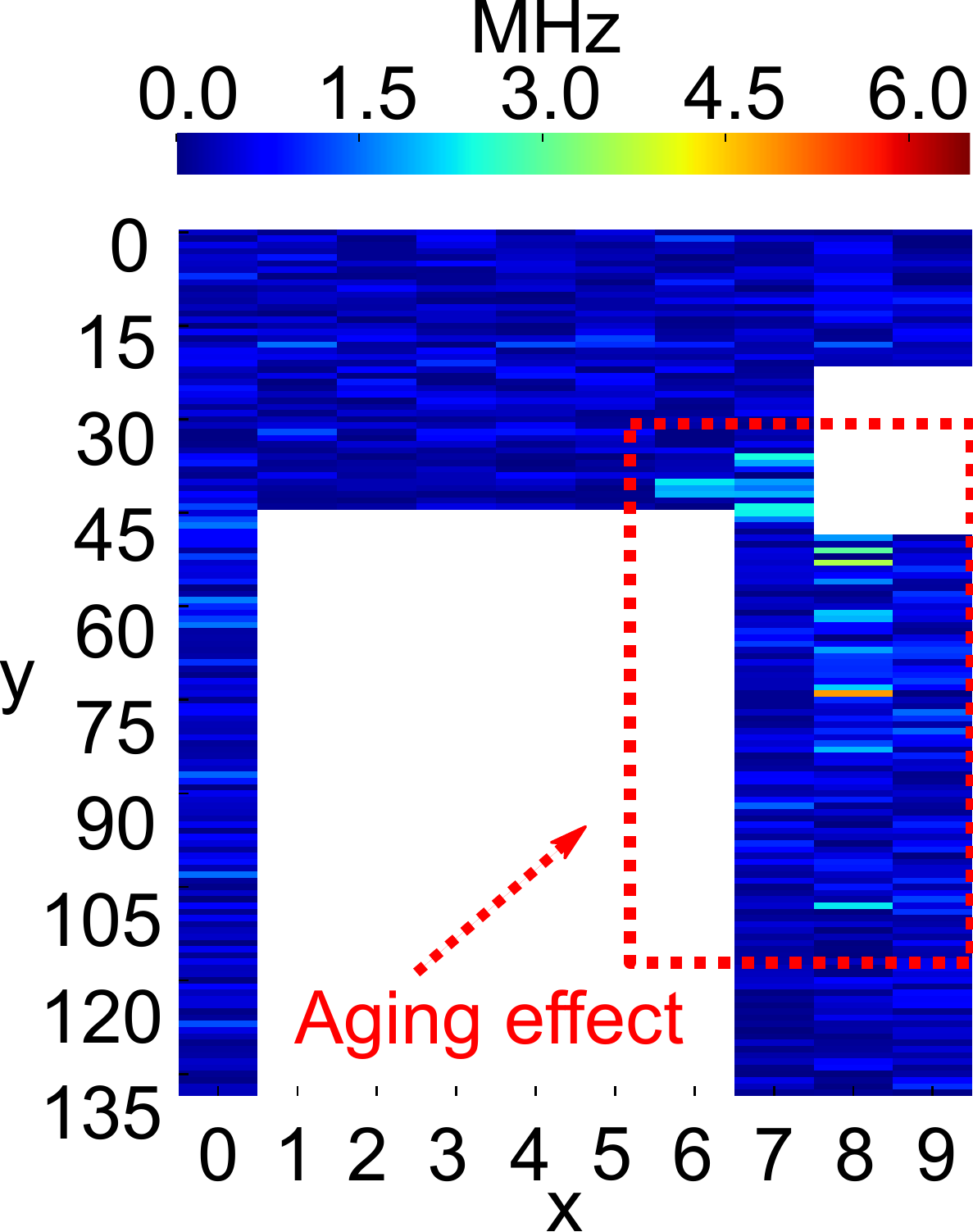}
    \label{fig:aged06_riscv}
  }
  \subfigure[\proof{3-h} aged FPGA-07]{
    \includegraphics[width=0.3\linewidth]{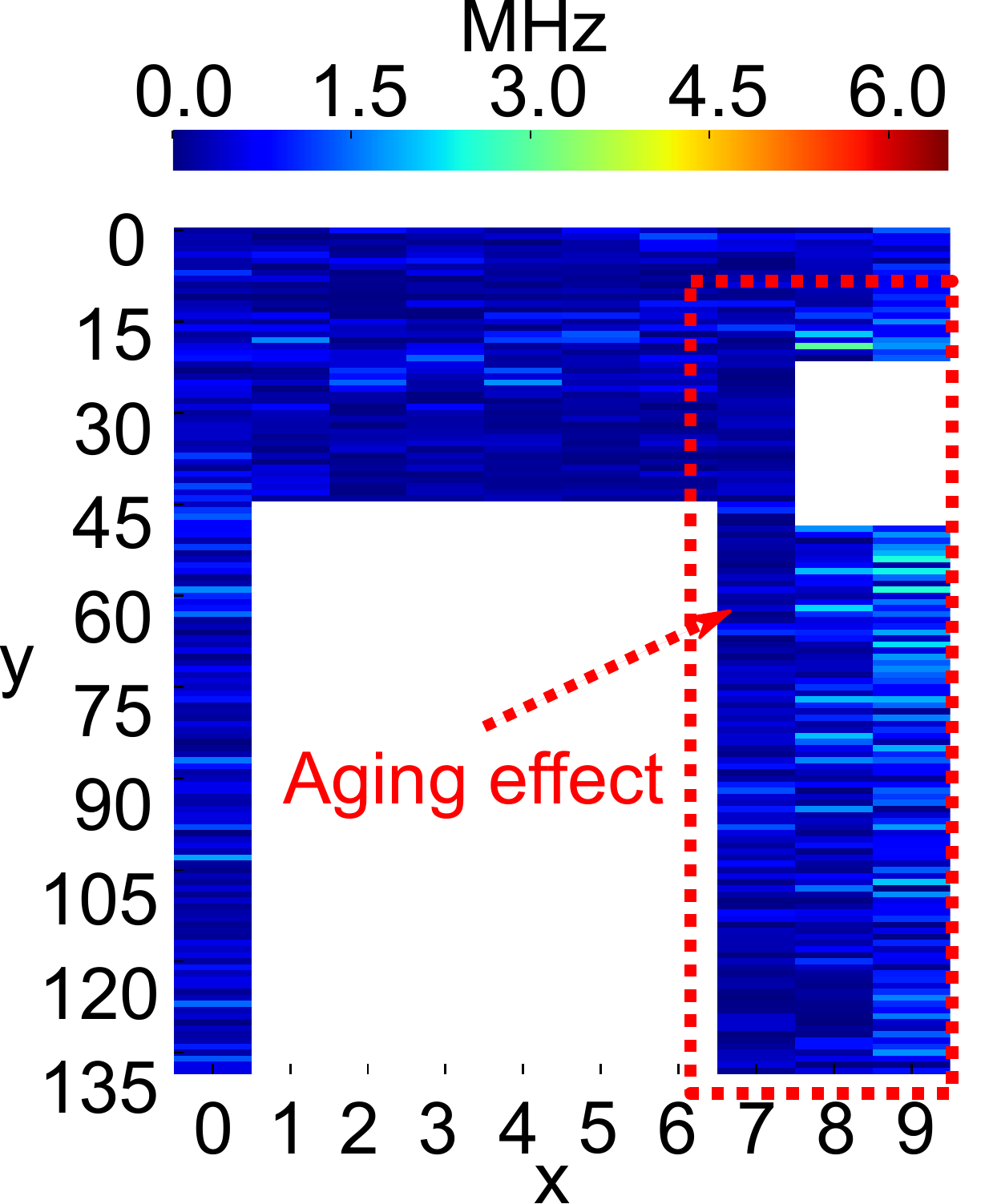}
    \label{fig:aged03_riscv}
  }\\
  \subfigure[\proof{2-h} aged FPGA-08]{
    \includegraphics[width=0.3\linewidth]{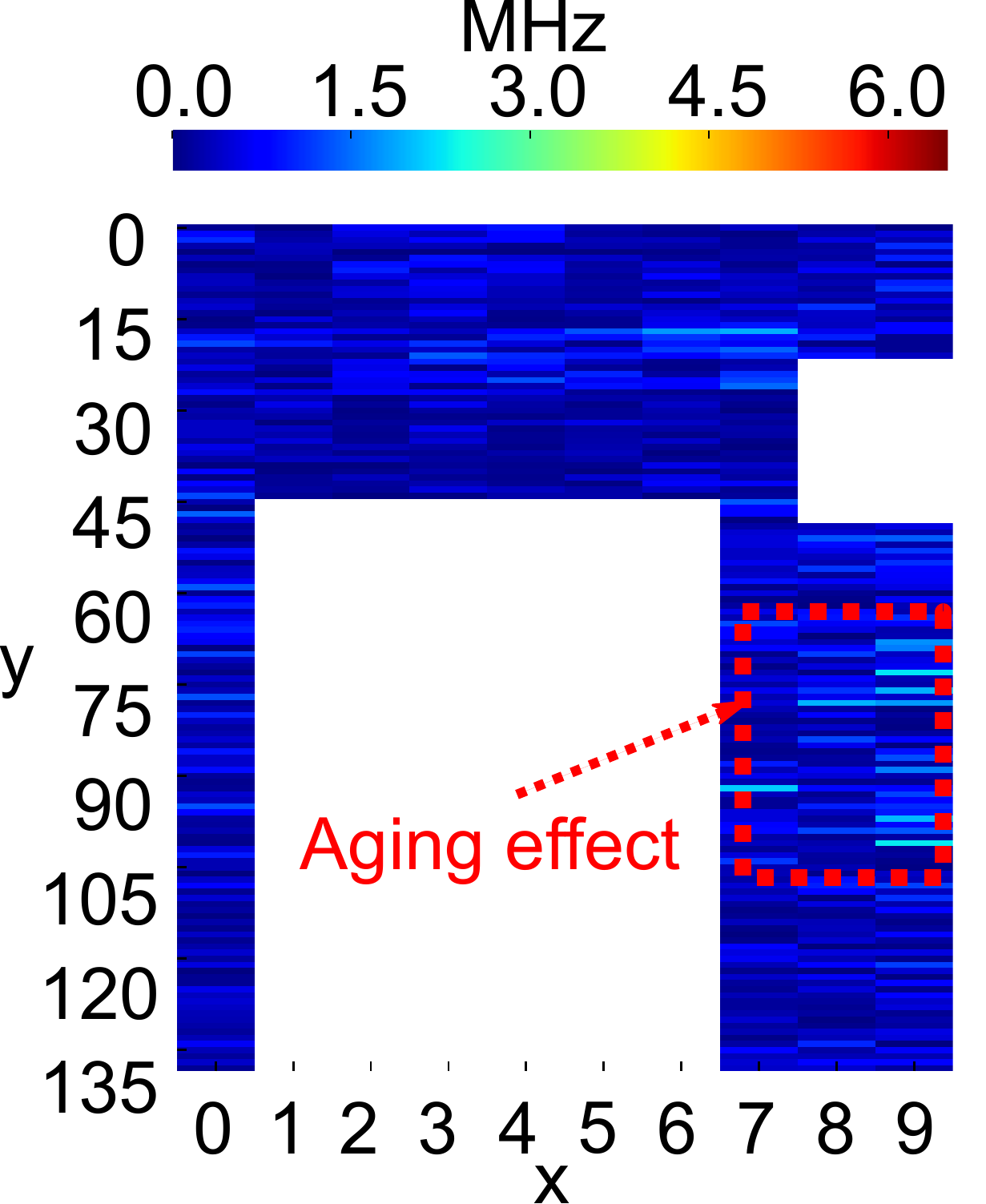}
    \label{fig:aged02_riscv}
  }
  \subfigure[\proof{1-h} aged FPGA-09]{
    \includegraphics[width=0.3\linewidth]{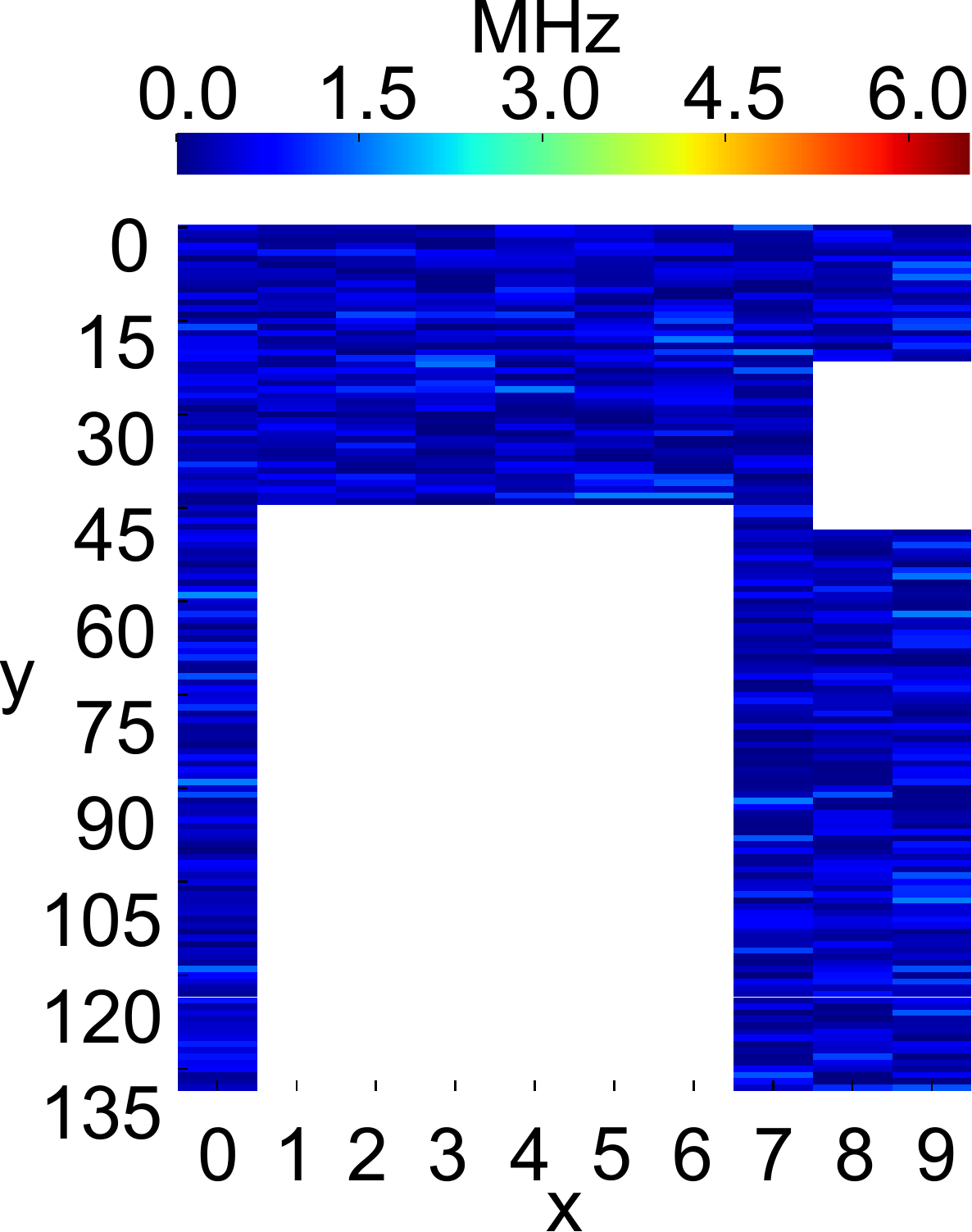}
    \label{fig:aged01_riscv}
  }
  \caption{Frequency heatmaps of the residuals of FPGA-06 to FPGA-09.}
  \label{fig:riscv_heatmap}
\end{figure}

In Fig.~\ref{fig:aged_residual}, the residual heatmaps of the aged
FPGA-01 to FPGA-05 are shown, where the paths with the highest anomaly scores in Fig.~\ref{fig:anomaly_riscv} are shown. \proof{From the figures},
the aging effect \proof{is} observed in the lower right area where
the s9234 benchmark circuit is located, \proof{for example}, a notable difference of
6\,MHz can be observed in Fig.~\ref{fig:aged01}. \proof{However}, in
Fig.~\ref{fig:aged05}, there is no significant difference in the
residuals between the fresh and the aged cases.
Fig.~\ref{fig:riscv_heatmap} also shows the frequency heatmaps of the
highest scores of the four aged FPGAs with the RISC-V processor
(FPGA-06 to FPGA-09).  From the figures, we can see \isaka{a} similar result
with Fig.~\ref{fig:aged_residual}.
\proof{Note} that the aging effect in the wider area was confirmed \proof{because}
the RISC-V processor has a larger circuit area than the s9234
circuit as shown in Figs.~\ref{fig:aged06_riscv}
and~\ref{fig:aged03_riscv}. This suggests that \proof{our} proposed
method works effectively even in large-scale circuits that
require \proof{the} full use of FPGA resources.

\subsubsection{Recycled FPGA detection}\label{sec:result}


\begin{figure}[!t]
  \centering
  \subfigure[Aged FPGA-01 to FPGA-05]{
    \includegraphics[width=0.58\linewidth]{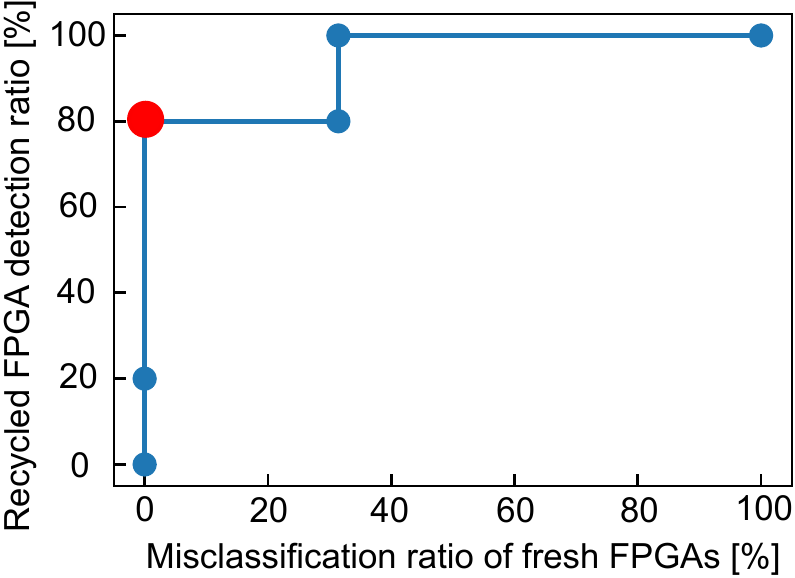}
    \label{fig:roc_s9234}
  }
  \subfigure[Aged FPGA-06 to FPGA-09]{
    \includegraphics[width=0.58\linewidth]{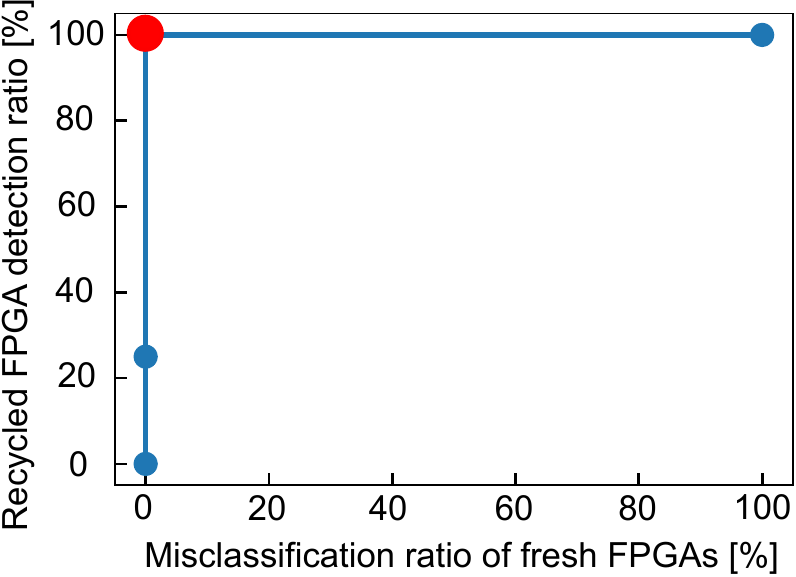}
    \label{fig:roc_riscv}
  }
  \caption{ROC curves of the proposed recycled FPGA detection method,
    where the 35 fresh FPGAs are used for both the cases. The
    highlighted points are the best performance.}
  \label{fig:roc}
\end{figure}

Finally, the \proof{accuracy of our proposed} detection \proof{method was} evaluated using the
anomaly scores.
The receiver operating characteristic (ROC) curve is \proof{represented by}
Fig.~\ref{fig:roc} \proof{and shows} the classification result of the
proposed method. In the ROC curves, the horizontal and vertical axes
express the misclassification result for the 35 fresh FPGAs and
correct classification result for the aged FPGAs, i.e., the upper left
corner of the ROC curves corresponds \isaka{to} the best result as highlighted in
Fig.~\ref{fig:roc}.

As shown in Fig.~\ref{fig:roc_s9234}, the proposed method can detect
the aged FPGAs with s9234 while failing to detect one aged FPGA.
\proof{The} proposed method detected the aged FPGAs with the
RISC-V processor without misclassification as shown in
Fig.~\ref{fig:roc_riscv}. \proof{In}
Fig.~\ref{fig:roc_s9234}, the undetected FPGA is the \proof{1-h} aged FPGA
(FPGA-05) as expected from the discussion of
Fig.~\ref{fig:1aged_s9234}.  Although it may be difficult to detect
the 1-h aged FPGA-08 in practical use, the FPGA can be detected
by setting a fresh/recycled threshold if the information on KFFs is
available as with~\cite{TVLSI2019_Alam}.  From these results, we
conclude that the proposed method detected a recycled FPGA regardless
of the user circuits when the FPGA was used for six days or more from
as represented in Eq.~(\ref{eq:thermal}).

For comparison, we \proof{applied} the recycled FPGA detection
using the
conventional method~\cite{TVLSI2019_Alam} as well as the proposed
method using the nine aged FPGAs (FPGA-01 to FPGA-09) and the 35 fresh
FPGAs (FPGA-01 to FPGA-35).  We randomly selected 265 RO
configurations; therefore, the evaluated frequency vector was composed
of 8,480 $(=265\times32$) frequencies.  We \proof{then} applied the $k$-means$++$
clustering algorithm to partition the frequencies into clusters. The
optimal number \proof{of clusters} was determined based on the silhouette
\proof{values obtained}. In~\cite{TVLSI2019_Alam}, it has been reported that the number
of the clusters is small when an FPGA is fresh, and the \isaka{number of clusters}
increases due to aging. Therefore, if the cluster number of
the FUT is larger than that of a fresh one, \proof{then the FPGA is}
recycled.

\begin{table}[!t]
    \centering
    \caption{
      \isakaNew{Silhouette value for each cluster in the case of the 265 randomly selected ROs.}
      The bold number is the maximum value, \proof{inferring to} the optimal \proof{cluster number}.}
    \label{tab:silhouette}
    \begin{tabular}{l|c|c|c}
      \hline
      & 2 cluster  &  3 cluster   & 4 cluster \\ \hline
      35 fresh FPGAs   & \textbf{0.557}  & 0.522  & 0.518 \\
      6-h aged FPGA-01  & \textbf{0.622}  & 0.613  & 0.592 \\ 
      6-h aged FPGA-02  & \textbf{0.625}  & 0.611  & 0.591 \\ 
      3-h aged FPGA-03  & \textbf{0.562}  & 0.549  & 0.548 \\ 
      2-h aged FPGA-04  & \textbf{0.545}  & 0.529  & 0.527 \\ 
      1-h aged FPGA-05  & \textbf{0.553}  & 0.525  & 0.520 \\ 
      6-h aged FPGA-06  & \textbf{0.611}  & 0.593  & 0.585 \\ 
      3-h aged FPGA-07  & \textbf{0.627}  & 0.614  & 0.580 \\ 
      2-h aged FPGA-08  & \textbf{0.554}  & 0.531  & 0.524 \\ 
      1-h aged FPGA-09  & \textbf{0.555}  & 0.531  & 0.523 \\ \hline
    \end{tabular}
\end{table}

Table~\ref{tab:silhouette} represents the silhouette values for each
FPGA when the number of clusters was 2, 3, and 4. From this table, the
numbers in bold represent the maximum silhouette value; that is, the
cluster number for the values in bold point to the optimal cluster
number. For the fresh FPGAs, the average of the silhouette values is
represented. In our study, the optimal \isaka{number of clusters} for
the fresh FPGAs was 2, while for all the aged FPGAs was also 2, with
no differences observed for the silhouette values. The result showed
the conventional method failed to identify the eight recycled FPGAs,
except for FPGA-06. We conclude that the RO selection was
insufficient for detection.

\begin{table}[!t]
    \centering
    \caption{\isakaNew{Silhouette value for each cluster when all the ROs are used.}
    \proof{The numbers in bold represent the maximum values obtained; pointing to the optimal number of clusters.}}
    \label{tab:silhouette_all}
    \begin{tabular}{l|c|c|c}
      \hline
      & 2 cluster  &  3 cluster   & 4 cluster \\ \hline
      35 fresh FPGAs   & \textbf{0.553}  & 0.523  & 0.522 \\
      6-h aged FPGA-01  & \textbf{0.556}  & 0.528  & 0.524 \\ 
      6-h aged FPGA-02  & \textbf{0.553}  & 0.526  & 0.524 \\ 
      3-h aged FPGA-03  & \textbf{0.555}  & 0.527  & 0.524 \\ 
      2-h aged FPGA-04  & \textbf{0.550}  & 0.525  & 0.522 \\ 
      1-h aged FPGA-05  & \textbf{0.552}  & 0.523  & 0.524 \\ 
      6-h aged FPGA-06  & \textbf{0.556}  & 0.527  & 0.525 \\ 
      3-h aged FPGA-07  & \textbf{0.554}  & 0.526  & 0.525 \\ 
      2-h aged FPGA-08  & \textbf{0.551}  & 0.529  & 0.519 \\ 
      1-h aged FPGA-09  & \textbf{0.550}  & 0.523  & 0.523 \\ \hline
    \end{tabular}
\end{table}

The detection performance of the conventional method depends on the
CLB selection.  To demonstrate the best performance \proof{by} the
conventional method, all the RO configurations were used for the
recycled FPGA detection. Table~\ref{tab:silhouette_all} shows the
silhouette values for each cluster as with
Table~\ref{tab:silhouette}. As a result, the optimal number of the
clusters remained 2 in all the cases.  We found it difficult to
distinguish between frequency fluctuations caused by the process
variation and those by aging degradation, nor did the clustering
algorithm work accurately.

In spite of demonstrating a good result in the paper
of~\cite{TVLSI2019_Alam}, the conventional method could not detect the
recycled FPGAs in our experiment. We consider this comes from the
difference in the experimental conditions. Although similar aging
acceleration equipment was used in the experiments of [10], the aging
schedule did not contain the recovery phase. Meanwhile, in our
experiment, the recovery phase of more than 4 days was added
considering a realistic scenario of recycled FPGAs in the
market. Thus, it can be considered that the recycled FPGA detection by
the conventional method was more difficult in our experiment.

\begin{table}[!t]
  \centering
  \caption{Comparison of the detection accuracy. The bold checkmarks
    express the correct results.}\label{tab:compare}
  \begin{tabular}{l|cc|cc|cc}
    \hline
    & \multicolumn{2}{c|}{Prop.} & \multicolumn{2}{c|}{Conv.} & \multicolumn{2}{c}{Conv.} \\
    & \multicolumn{2}{c|}{} & \multicolumn{2}{c|}{w/ 265 ROs} & \multicolumn{2}{c}{w/ all ROs} \\
    & Fre.     & Recy.       & Fre.         & Recy.      & Fre.         & Recy.    \\ \hline
    35 fresh FPGAs    & \CheckmarkBold         &                &  \CheckmarkBold             &             &  \CheckmarkBold             &             \\
    6-h aged FPGA-01 &           & \CheckmarkBold              & \checkmark             &    &  \checkmark             &             \\
    6-h aged FPGA-02 &           & \CheckmarkBold              & \checkmark             &    &  \checkmark             &             \\
    3-h aged FPGA-03 &           & \CheckmarkBold              & \checkmark             &    &  \checkmark             &             \\
    2-h aged FPGA-04 &           & \CheckmarkBold              & \checkmark             &    &  \checkmark             &             \\
    1-h aged FPGA-05 & \checkmark         &                & \checkmark             &             &  \checkmark             &             \\
    6-h aged FPGA-06 &           & \CheckmarkBold              & \checkmark             &       &  \checkmark             &             \\
    3-h aged FPGA-07 &           & \CheckmarkBold              & \checkmark             &             &  \checkmark             &             \\
    2-h aged FPGA-08 &           & \CheckmarkBold              & \checkmark             &             &  \checkmark             &             \\
    1-h aged FPGA-09 &           & \CheckmarkBold              & \checkmark             &             &  \checkmark             &             \\ \hline
  \end{tabular}
\end{table}

As a summary of the comparison of the proposed method and conventional
method, the detection accuracy of the recycled FPGA is shown in
Table~\ref{tab:compare}. In this table, the best performances
highlighted in Fig.~\ref{fig:roc} are shown as the result of the
proposed method. The proposed and conventional methods could identify
all the 35 fresh FPGAs as fresh. The proposed method classified the
eight aged FPGAs but missed the 1-hour aged FPGA (FPGA-05) as
recycled. However, the conventional method failed to detect all the
aged FPGAs even when all the ROs are used. From these results, we can
conclude that the proposed method can classify the recycled FPGAs
better than the conventional method in the experiment of degradation
using the s9234 circuit and RISC-V processor.

\section{Conclusion}\label{sec:conclusion}
In this study, we proposed an accurate KFF-free recycled FPGA
detection method in which \proof{we mitigated} the process variation
effect in the measured RO \proof{frequencies.} \proof{We reduced} the
systematic component \proof{of the RO frequency} by comparing the RO
frequencies of neighboring CLBs, which also \proof{contribute} to
eliminating the need for a set of KFFs.  In addition, the random
component of the RO frequency \proof{was} canceled out by increasing
the number of RO stages to further improve the detection
accuracy. Recycled FPGAs \proof{were} effectively detected using an
unsupervised outlier detection method based on direct density ratio
estimation.  Experimental results using 35 commercially available
FPGAs demonstrated that the proposed method successfully mitigates
process variation effect and captures aging effect to detect recycled
FPGAs. \proof{We further} confirmed that the proposed method
successfully detected all the recycled FPGAs except for that with a
short aging time, while the conventional method failed to detect aged
FPGAs.

As future work, we intend to further improve the
detection performance of unsupervised recycled FPGA detection. As
discussed in Section~\ref{sec:exp}, the proposed method failed to
detect FPGAs with a short aging time.
It is expected that it could be improved by
introducing the degradation analysis proposed in~\cite{TCAD_KeHuang}
into the unsupervised detection framework with density ratio
estimation of the proposed method.

\ifCLASSOPTIONcaptionsoff
  \newpage
\fi

\bibliographystyle{IEEEtran}
\bibliography{alias,ref}

%








\end{document}